\documentclass[reprint,amsmath, amssymb, aps, prl, citeautoscript, superscriptaddress, nobibnotes, longbibliography]{revtex4-2}

\usepackage{multirow}
\usepackage{minitoc}
\usepackage{appendix}
\usepackage{enumitem}
\usepackage{booktabs}
\usepackage{float}

\usepackage{soul}
\usepackage{siunitx}
\usepackage{graphicx}
\usepackage{braket}
\usepackage[colorlinks]{hyperref}
\usepackage[capitalize]{cleveref}
\usepackage[normalem]{ulem}
\usepackage{comment}

\hypersetup{citecolor={black}}  

\newcommand{\sivtn}{$^{29}$SiV}

\makeatletter
\newcommand*{\balancecolsandclearpage}{%
  \close@column@grid
  \newpage
  \twocolumngrid
}
\makeatother

\makeatletter
\newcommand*{\balancecolsandclearpagesingle}{%
  \close@column@grid
  \newpage
}
\makeatother

\begin{document}

\preprint{APS/123-QED}

\title{Entanglement Assisted Non-local Optical Interferometry in a Quantum Network}

\author{P.-J.~Stas}
 \thanks{These authors contributed equally to this work.}
 \affiliation{Department of Physics, Harvard University, Cambridge, Massachusetts 02138, USA}
\author{Y.-C.~Wei}
\thanks{These authors contributed equally to this work.}
 \affiliation{Department of Physics, Harvard University, Cambridge, Massachusetts 02138, USA}
\author{M.~Sirotin}
\thanks{These authors contributed equally to this work.}
\affiliation{Department of Physics, Harvard University, Cambridge, Massachusetts 02138, USA}
 \affiliation{Department of Physics and Research Laboratory of Electronics, Massachusetts Institute of Technology, Cambridge, Massachusetts 02138, USA}
\author{Y.~Q.~Huan}
 \affiliation{Department of Physics, Harvard University, Cambridge, Massachusetts 02138, USA}
\author{U.~Yazlar}
\affiliation{Department of Physics, Harvard University, Cambridge, Massachusetts 02138, USA}
 \affiliation{Division of Materials Science \& Engineering, Boston University, Boston, Massachusetts 02215, USA}
 \author{F.~Abdo~Arias}
 \affiliation{Department of Physics, Harvard University, Cambridge, Massachusetts 02138, USA}
 \author{E.~Knyazev}
 \affiliation{Department of Physics, Harvard University, Cambridge, Massachusetts 02138, USA}
 \author{G.~Baranes}
 \affiliation{Department of Physics, Harvard University, Cambridge, Massachusetts 02138, USA}
 \affiliation{Department of Physics and Research Laboratory of Electronics, Massachusetts Institute of Technology, Cambridge, Massachusetts 02138, USA}
\author{B.~Machielse}
 \affiliation{John A. Paulson School of Engineering and Applied Sciences, Harvard University, Cambridge, Massachusetts 02138, USA}
 \affiliation{IonQ Inc., 4505 Campus Dr., College Park, Maryland 20740, USA}
 \author{S.~Grandi}
 \affiliation{ICFO-Institut de Ciencies Fotoniques, The Barcelona Institute of Science and Technology, 08860 Castelldefels (Barcelona), Spain}
 \author{D.~Riedel}
 \affiliation{IonQ Inc., 4505 Campus Dr., College Park, Maryland 20740, USA}
 \author{J.~Borregaard}
 \affiliation{Department of Physics, Harvard University, Cambridge, Massachusetts 02138, USA}
\author{H.~Park}
 \affiliation{Department of Physics, Harvard University, Cambridge, Massachusetts 02138, USA}
 \affiliation{Department of Chemistry and Chemical Biology, Harvard University,
 Cambridge, Massachusetts 02138, USA}
\author{M.~Lon{\v{c}}ar}
 \affiliation{John A. Paulson School of Engineering and Applied Sciences, Harvard University, Cambridge, Massachusetts 02138, USA}
\author{A.~Suleymanzade}
 \affiliation{Department of Physics, Harvard University, Cambridge, Massachusetts 02138, USA}
 \affiliation{Department of Physics, University of California, Berkeley, California 7300, USA}
\author{M.~D.~Lukin}
 \altaffiliation{Corresponding author. E-mail:  lukin@physics.harvard.edu.}
 \affiliation{Department of Physics, Harvard University, Cambridge, Massachusetts 02138, USA}

\date{\today}

\begin{abstract}
The sensitivity of non-local optical measurements at low light intensities, such as those involved in long baseline telescope arrays, can be improved by using remote entanglement. Here, we demonstrate the use of entangled quantum memories in a quantum network of Silicon-vacancy centers in diamond nano-cavities to experimentally perform such non-local phase measurements. Specifically, we combine the generation of event-ready remote quantum entanglement, photon mode erasure that hides the ``which-path'' information of temporally and spatially separated incoming optical modes, and non-local, non-destructive photon heralding enabled by remote entanglement to  perform a proof-of-concept entanglement-assisted differential phase measurement of weak incident light between two spatially separate stations. Demonstrating successful operation of the remote phase sensing protocol with a fiber link baseline up to 1.55 km, our results open the door for a new class of quantum enhanced imaging applications. 
\end{abstract}

\maketitle
Optical interferometry is a well-established method for high-resolution imaging with wide-ranging applications from physics and astronomy to biological and medical imaging \cite{Hariharan_2012, Baldwin_Haniff_2002, Nolte_2012}. For instance, astronomical interferometry is routinely used for the observation of stellar objects where the light signal from multiple physically separated telescopes is combined to increase the imaging resolution \cite{VLA, NPOI, CHARA, EHTC}. In such a case, an array of optical receivers forms a synthetic aperture whose resolution scales with their separation (the baseline) \cite{Baldwin_Haniff_2002}. However, increasing the baseline of receiver arrays in practice is challenging \cite{Darre2016}. In the limit of weak signals typical of the optical domain, the optimal method for observation is direct interference of incident photons \cite{Tsang2011}, which is hindered by the exponential loss of signal light associated with optical fiber-based connections \cite{Azuma2023}. Quantum networks \cite{KimbleQuantumInternet, WehnerQuantumInternet} provide a way to perform non-local interference measurements. The key idea is to use quantum entanglement to effectively teleport the quantum state between remote receiver nodes (Fig.~\ref{fig1}a right panel), thereby enabling direct interference \cite{gottesman2012, Zhang2025}. While a scheme involving entanglement for non-local interference has been recently demonstrated in an all-photonic setting \cite{Brown2023}, the use of quantum memories presents a practical path towards overcoming photon loss through event-ready (heralded) entanglement
 \cite{gottesman2012} and efficient information processing together with local photon mode erasure, enabling an exponential reduction in the number of required entangled pairs \cite{emil_telescopes, emil_PRA}. 

Here, we demonstrate quantum memory-assisted non-local interferometry with a two-station network separated by a line-of-sight distance of $\sim$ 6 meters (Fig.~\ref{fig1}b,c). Our approach employs atom-like defects in solid state \cite{Pompili2021, Guo2023, Riedel2023, Ngan2023, Stolk2024, Photonic2024, Ruskuc2025, Harris2025}, particularly Silicon-vacancy centers (SiV) integrated in diamond nanophotonic cavities. Such systems recently emerged as a promising platform for quantum networking due to their access to long-lived spin quantum memories, high gate and readout fidelities, and strong light-matter interaction enabling efficient spin-photon operations \cite{Nguyen2019PRL, si29_node}. These properties have enabled experimental implementations of quantum memory-enhanced communication \cite{Bhaskar2020}, entanglement generation over a metropolitan-scale deployed fiber \cite{two_fridge}, and blind quantum computation \cite{BQC_SiV}. In our experiments, each SiV constitutes a two-qubit register with a communication qubit (electron spin) and a memory qubit ($^{29}$Si nuclear spin). Signal photons are reflected off the fiber-coupled SiV-cavity systems and an optical fiber network is used for readout, entanglement generation, and signal light collection \cite{Nguyen2019PRL, si29_node}. We utilize an improved parallel instead of serial \cite{two_fridge} entanglement scheme to reach higher entanglement rate, and demonstrate non-destructive photon heralding, both locally with a time-bin photonic qubit on a single station, as well as non-locally for a photon in superposition between two remote spatial modes by using remote entanglement. Such photon heralding filters out vacuum fluctuations to achieve optimal interferometer sensitivity~\cite{Tsang2011}. We combine this method with photon mode erasure by interfering the arriving photons at two nodes with a coherent state of light to hide the which-path-information for interferometric measurements. Finally, we integrate these elements to demonstrate the operation of a long-baseline quantum memory-assisted interferometer with a fiber separation length of up to 1.55~km, five times larger than the current state-of-the art optical telescope array baseline of 330~m \cite{CHARA}.

\section{Non-local phase sensing}

\begin{figure*}
    \centering
    \includegraphics[width=\textwidth]{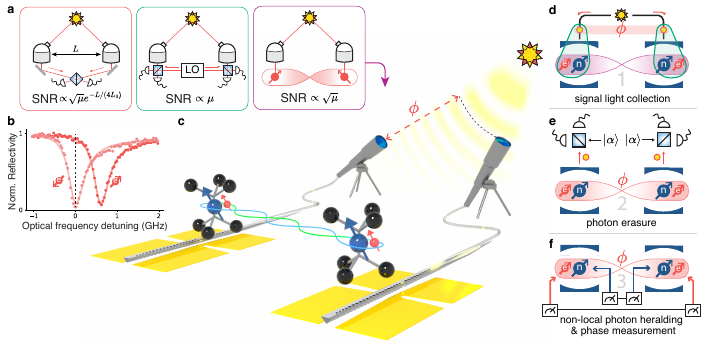}
    \caption{\textbf{A quantum memory-assisted non-local interferometer based on a quantum network.} a)~Three remote phase sensing methods and their signal-to-noise ratio (SNR) scalings from left to right: direct interferometric detection, local measurements with a local oscillator (LO), and entanglement-assisted non-local phase sensing. b)~The electron spin state-dependent cavity reflectance
 near the SiV optical transition is used for signal photon storage and quantum operations. The dashed line indicates the frequency used for electron spin state readout, photonic entanglement, and signal light collection. c)~Entangled qubits shared between the stations are used to improve the sensitivity of a non-local interferometer. Once entanglement has been generated, the steps of the quantum memory-assisted remote phase sensing protocol are d)~signal light collection through local operations, e)~signal photon mode erasure to complete photon state storage, and f)~non-local photon heralding through electron state measurement, and phase probing through nuclear state measurement.}
    \label{fig1}
\end{figure*}

The signal in a non-local interferometer, such as the angle of the incident light from a distant object at two detector stations, is proportional to the differential phase $\phi$ between the detector stations (Fig.~\ref{fig1}c). The aim of the imaging is thus to efficiently and precisely measure this interferometric phase $\phi$. In conventional systems, two approaches to measure $\phi$ can be distinguished for thermal light. The first involves direct interference of non-locally collected photons (Fig.~\ref{fig1}a left panel), while the second involves local measurements (Fig.~\ref{fig1}a middle panel). In the first approach direct interference of the signal light collected from each station is enabled by routing the light to a central beamsplitter. This method achieves optimal interference visibility with signal-to-noise ratio (SNR) scaling $\sim \sqrt{\mu_{\mathrm{sig}}}$ \cite{Tsang2011}, where $\mu_{\mathrm{sig}} \ll 1$ is the average photon number of the incident light. However, for long baseline measurements, this approach 
introduces signal photon loss that typically scales exponentially with the distance between the stations \cite{Azuma2023}. The second approach consists of interfering the collected photons with a distributed local oscillator (LO) at each station. The phase difference between the two stations can then be determined by classically comparing the local measurement results. However, as the signal light is mixed with the LO, the local measurements cannot distinguish the events when a signal photon has arrived from those when it has not. Measurements without signal photons (corresponding to vacuum input) carry no useful phase information but introduce shot noise (vacuum fluctuation noise), reducing the interference visibility and resulting in a SNR scaling as $\sim\mu_{\mathrm{sig}}$ \cite{Tsang2011}. While it is also possible to independently measure the local phase of the incident light at each station using higher-order correlations \cite{HBT_1957, Glauber1963}, 
for thermal light this requires the simultaneous arrival of a signal photon at each station, which maintains the unfavorable scaling \cite{Tsang2011}.

Entangled quantum memories provide a route to achieve optimal non-local measurements without the exponential photon loss with the baseline size~\cite{emil_telescopes, gottesman2012} (Fig.~\ref{fig1}a right panel). Specifically, pre-generated entanglement between the stations can be utilized as a resource to perform non-local photon heralding in which the arrival of a signal photon can be detected without revealing at which station it arrives~\cite{emil_telescopes}. This allows one to distinguish vacuum from signal photons without destroying the phase information $\phi$. By keeping only measurement results with successful non-local heralding, vacuum fluctuations can be effectively filtered out to increase the visibility and SNR (see Methods for details).

To realize this method experimentally, our approach consists of first ``arming" the interferometer by preparing the nuclear qubits in an entangled state between two stations. The entanglement is event-ready as it is heralded, a key improvement over all-photonic approaches \cite{Brown2023}. We model the distributed signal light with a weak laser pulse with average photon number $\mu_{\mathrm{sig}} \ll 1$:
$\ket{\psi_{\mathrm{sig}}} \sim \ket{0_L0_R} + \sqrt{\mu_{\mathrm{sig}}/2}(\ket{0_L1_R} + e^{i\phi}\ket{1_L0_R})$,
where $\ket{0_L0_R}$ corresponds to vacuum, and $\ket{1_L0_R}$ ($\ket{0_L1_R}$) corresponds to a single photon arriving at the left (right) station. The local signal phase at each station is averaged over a uniform distribution (with fixed differential phase $\phi$), such that the signal effectively behaves like a thermal state in the weak signal regime \cite{Tsang2011} (see Methods, Extended Data Fig. \ref{SIfig_phase_dist}). The photonic signal is collected through local quantum operations (Fig.~\ref{fig1}d) that entangle the photonic state with the qubits at each station. We then erase the photonic mode information (Fig.~\ref{fig1}e) \cite{emil_PRA} and subsequently implement non-local, non-destructive photon heralding by measuring the parity of electron qubit spins at the two station. Such a correlated parity measurement heralds the arrival of a photon without revealing which station the photon arrived at (Fig.~\ref{fig1}f). Finally, the differential phase $\phi$ of the photon imprinted on the initial Bell state between the nuclear spins is obtained through a 2-qubit nuclear parity measurement performed locally. 

\section{Parallel entanglement generation}

\begin{figure}
    \centering
    \includegraphics[width=\linewidth]{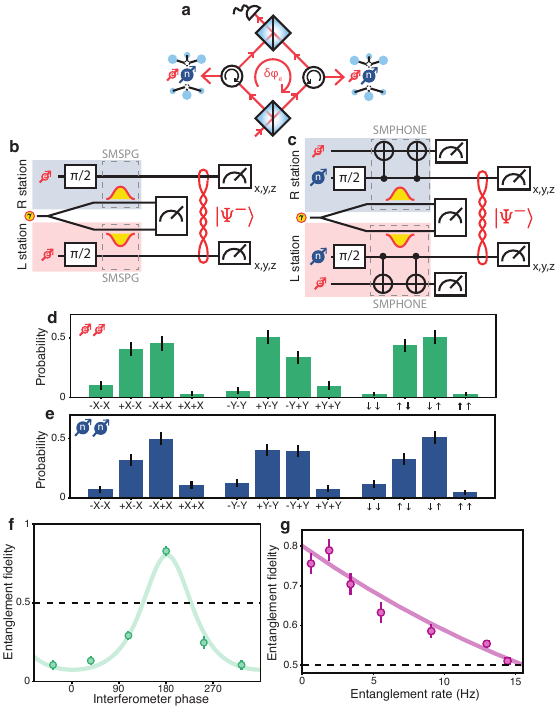}
    \caption{\textbf{Parallel entanglement generation between SiV centers} a)~SiV stations connected in a Mach-Zehnder interferometer setup. b)~Electron-electron entanglement circuit using the SMSPG. c)~Nucleus-nucleus entanglement circuit using the SMPHONE. For both b) and c), all rotations are around the $y$-axis. Bell state tomography for d)~electron-electron and e)~nucleus-nucleus entanglement. f)~Electron-electron entanglement fidelity as a function of interferometer phase lock point. g)~Electron-electron entanglement fidelity as a function of entanglement rate, sweeping the average photon number in the weak entanglement laser pulse from 0.1 to 1. Error bars in d), e), f) and g) are 1 s.d.}
    \label{fig2}
\end{figure}

In prior work, SiV remote entanglement generation has relied on serial entangling schemes \cite{two_fridge, BQC_SiV}. Here we implement a parallel entangling scheme \cite{Beukers2025, Levonian2022} with a 7.5 times higher efficiency
\cite{two_fridge}. This is realized by connecting the two stations in a Mach-Zehnder interferometer configuration that must be phase-stable with each path reflecting off one SiV-cavity system (Fig.~\ref{fig2}a). In this approach, we generate entanglement between electron spin qubits by splitting a weak laser pulse on a beamsplitter to send a dual-rail photonic qubit in superposition to each station with electron spins initially in $\ket{+}$ state. We then perform a single-mode spin-photon gate (SMSPG, Fig.~\ref{fig2}b), which relies on the spin state-dependent conditional reflection amplitude between the photon and the electron spin qubit $\ket{1_{\gamma}+_e}\rightarrow\ket{1_{\gamma}\uparrow_e}/\sqrt{2}$ at each station (Fig.~\ref{fig1}b). We note that this gate is non-unitary due to photon loss (see Supplementary Information), corresponding to the transformation:
\begin{multline}
    (\ket{0_L1_R} + \ket{1_L0_R})/\sqrt{2}\ket{+_{e_L}+_{e_R}} \\
    \longrightarrow (\ket{0_L1_R}\ket{+_{e_L}\uparrow_{e_R}} + \ket{1_L0_R}\ket{\uparrow_{e_L} +_{e_R}})/\sqrt{2}.
\end{multline}
The dual-rail photon is then recombined at a beam splitter that projects the photon in the $\ket{01} \pm e^{i\delta\phi_e}\ket{10}$ basis. This step heralds a successful entanglement attempt, ensuring that the fidelity does not degrade at low success probabilities. For measurement outcome $\ket{01} + e^{i\delta\phi_e}\ket{10}$ the resulting electrons' state (up to normalization) is:
\begin{equation}
\label{eq:final_ent_state}
    (1+e^{i\delta\phi_e})(\ket{\uparrow\uparrow} + \ket{\Psi^+}/2) + (1-e^{i\delta\phi_e})\ket{\Psi^-}/2,
\end{equation}
where $\ket{\Psi^{\pm}}$ corresponds to the entangled Bell state $\ket{\uparrow\downarrow}\pm\ket{\downarrow\uparrow}$. By locking the interferometer phase $\delta\phi_e$ to $\pi$, we prepare the Bell state $\ket{\Psi^-}$ across the two stations (Fig.~\ref{fig2}d,f) with a fidelity of $F=0.83(3)$.  
By tuning the average photon number of the incident light used in the entanglement protocol, we can increase the entanglement success probability per trial at the cost of reduced entangled state fidelity due to multi-photon contributions. We reach an entanglement rate of 13 Hz for $F\geq 0.5$ and 1.9~Hz for $F = 0.79(3)$ (Fig.~\ref{fig2}g), enabling practical entanglement-based sensing experiments.

The non-local phase measurement protocol requires entanglement between the nuclear spins, which we generate by replacing the SMSPG with single-mode photon-nucleus entangling gates (SMPHONE, Fig.~\ref{fig2}c), which generates the state $\ket{\uparrow_{e_L}\uparrow_{e_R}}\ket{\Psi^-}_{n_L, n_R}$ (Fig.~\ref{fig2}e). Similar to the PHONE gate \cite{si29_node, two_fridge}, the electron ends in the $\ket{\uparrow}$ state unless a gate error occurred. By reading out the electron state we can detect errors and discard electron $\ket{\downarrow}$ measurements, reaching a nuclear Bell state fidelity of $F=0.73(4)$.

\section{Photon erasure and non-local heralding}

\begin{figure}
    \centering
    \includegraphics[width=\linewidth]{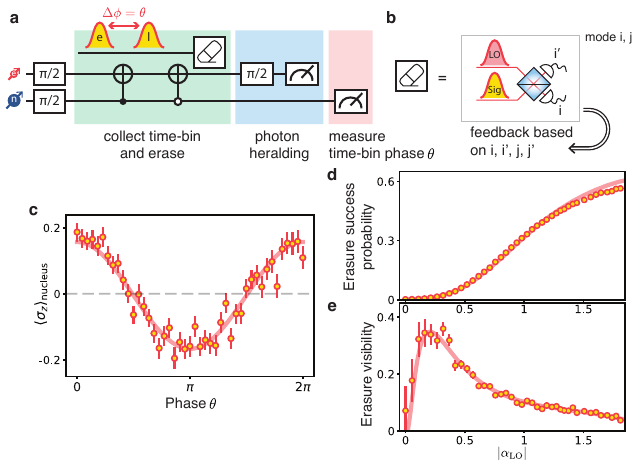}
    \caption{\textbf{Erasing and heralding photons in two temporal modes on a single node.} a)~Experimental sequence to measure phase difference between two temporal modes of a photon using erasure and heralding, with all rotations around the $y$-axis. For the C$_n$NOT$_e$ gates, the black (white) circle gate conditional on the nucleus being in state $\ket{\downarrow}$ ($\ket{\uparrow}$) b)~Photon erasure protocol for any two photonic modes (temporal, spatial, etc.). c)~Nuclear state measurement result after the sequence shown in a). d)~Success probability and e)~signal visibility of sequence a) as a function of coherent state strength of LO. Error bars in c), d) and e) are 1 s.d.}
    \label{fig3}
\end{figure}

Photon mode erasure \cite{emil_PRA} is a crucial step in our protocol to store the signal phase information onto the quantum memories while being compatible with efficient signal storage techniques \cite{emil_telescopes}. After the interaction with the SiV at each station, we must ensure that the which-path information (i.e., at which station the photon has arrived) is not extracted to preserve the differential signal phase information ($\phi$). 
 We achieve this by interfering the photonic signal with an LO coherent state on a beam splitter at each station and measuring the output ports with photon number resolving detectors (Fig.~\ref{fig3}b). 
We apply feedback on the nuclear qubits depending on the number of clicks at the first $i$ ($j$) and second $i'$ ($j'$) port of the left (right) station mode, and post-select on click outcomes $(i\neq i')\& (j\neq j')$ (see Supplementary Information) \cite{emil_PRA}.
Because the LO contains vacuum, single-photon, and multi-photon components, detection cannot distinguish whether the photons originated from the LO or from the signal. To identify the photon arrival, we then utilize the electron spins to herald the presence of a signal photon---without revealing in which mode this photon is located---and imprint its phase on the nuclear spins.

As a first step towards realizing this protocol, we demonstrate the local implementation of erasure and heralding of a photon in a superposition of two temporal modes ($\ket{1_e0_l} + \ket{0_e1_l}$, with $\ket{i_{e(l)}}$ the photonic state of the early (late) temporal mode, separated by 1 $\mu$s). Since in this case the photon heralding is local, it does not require a Bell pair. Instead we use an electron $X$-basis measurement to herald the arrival of a signal photon while hiding the temporal information. We note that this experimental sequence can be applied for the implementation of quantum receivers for classical communication \cite{Smith2025}.

\begin{figure*}
    \centering
    \includegraphics[width=0.8\textwidth]{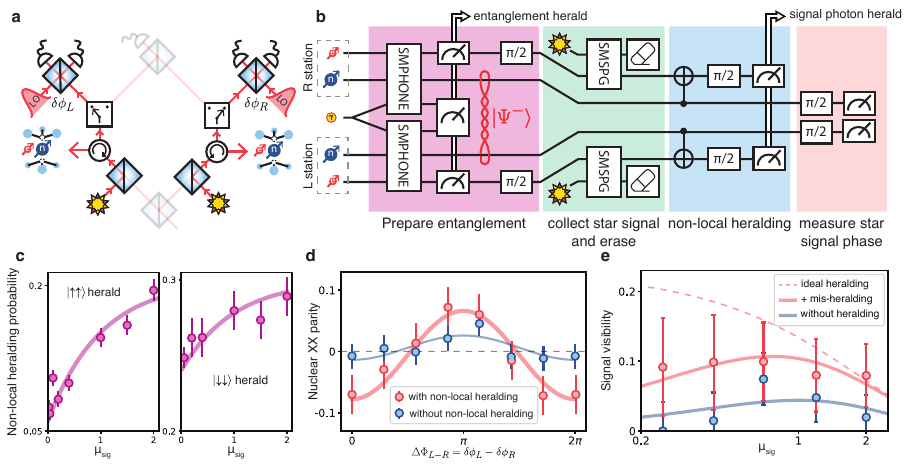}
    \caption{\textbf{Implementation of quantum memory-assisted interferometry.}  a)~Signal light collection and routing to photon erasure. The shaded components are those used in the entanglement generation. b)~Circuit diagram of the non-local phase sensing protocol implementation. All rotations are around the $y$-axis. c)~Left (right): Non-local signal photon heralding probability on the electron $\ket{\uparrow_e\uparrow_e}$ ($\ket{\downarrow_e\downarrow_e}$) state as a function of the average signal photon number. d)~Nuclear 2-qubit XX parity expectation value as a function of station differential phase $\Delta\Phi_{L-R}$ (including LO local phases, see Methods) with (red) and without (blue) non-local signal photon heralding, averaged over all $\mu_{\mathrm{sig}}={0.25,...,2}$. e)~Nuclear parity measurement visibility as a function of average signal photon number $\mu_{\mathrm{sig}}$ arriving at the SiVs with (red) and without (blue) non-local signal photon heralding. The solid blue (dashed red) curve corresponds to the visibility scaling without (with ideal) heralding, and the solid red curve corresponds to imperfect heralding due to mis-heralding (see Methods for more details). Error bars in c), d) and e) are 1 s.d.}
    \label{fig4}
\end{figure*}

In our experiment, starting with the photon state $\ket{0_e0_l} + \sqrt{\mu_{\mathrm{sig}}/2}(\ket{1_e0_l} + e^{i\theta}\ket{0_e1_l})$, we implement the gate sequence shown in Fig.~\ref{fig3}a resulting in the state:
\begin{multline}
    \ket{0_e0_l}\ket{\downarrow}_{elec}\ket{+}_n + \sqrt{\mu_{\mathrm{sig}}/2}[\ket{1_e0_l}\ket{+}_{elec}\ket{+}_n \\+ e^{i\theta}\ket{0_e1_l}(\ket{+}_{elec}\ket{\downarrow}_n + \ket{-}_{elec}\ket{\uparrow}_n)/\sqrt{2}].
\end{multline}
We then perform the photon erasure with feedback $X^{(1-\mathrm{sign}[(i-i')(j-j')])/2}$
on the nucleus, which is equivalent to measuring each temporal mode in the $X$-basis. This results in the state:
\begin{multline}
    \ket{\downarrow}_{elec}\ket{+}_n + \sqrt{\mu/2}[\ket{+}_{elec}\ket{+}_n \\
    + e^{i\theta}(\ket{+}_{elec}\ket{\downarrow}_n + \ket{-}_{elec}\ket{\uparrow}_n)/\sqrt{2}],
\end{multline}
effectively realizing a one-bit teleportation \cite{zhou2000} such that the signal photon differential phase remains imprinted on the nuclear spin state. By measuring the electron spin state and selecting the events with outcome $\ket{\uparrow}_{elec}$ only, we then non-destructively herald the presence of a signal photon, regardless of whether it was in the early or late temporal mode (Extended Data Fig.~\ref{SIfig_timebin_heralding}), yielding the nuclear state:
\begin{equation}
    (1+e^{i\theta})/2\ket{\downarrow}_n + (1-e^{i\theta})/2\ket{\uparrow}_n.
\end{equation}
Experimentally, we determine $\theta$ by measuring the nucleus in the $Z$-basis (Fig.~\ref{fig3}c). Increasing the strength $|\alpha_{\mathrm{LO}}|=\sqrt{\mu_{\mathrm{LO}}}$ of the LO pulses in the erasure increases the efficiency (Fig.~\ref{fig3}d) and reaches a maximum visibility of 0.36(3) at $\alpha_{\mathrm{LO}}=0.32$ (Fig.~\ref{fig3}e). The visibility is limited by photon loss (including the spin-photon gate efficiency) between the SiV and the final photon detectors at large $|\alpha_{\mathrm{LO}}|$ and detector dark counts at small $|\alpha_{\mathrm{LO}}|$, as well as MW errors (Extended Data Tab. \ref{SItable:vis_budget}, see Supplementary Information).

\section{Quantum memory-assisted interferometry implementation}

\begin{figure}
    \centering
    \includegraphics[width=0.65\linewidth]{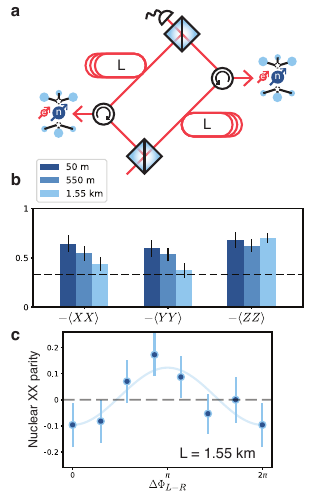}
    \caption{\textbf{Extending the baseline of the interferometer to 1.55 km} a)~Additional spools of length $L$ added in the entanglement interferometer. A baseline length of $L$ requires a fiber length of $2L$ within the entanglement interferometer. b)~Nucleus-nucleus entanglement Bell state tomography for different baseline lengths. The dashed line shows the classical limit of $1/3$. c)~Nuclear 2-qubit $XX$ parity expectation value as a function of station differential phase $\Delta\Phi_{L-R}$ for a baseline of 1.55 km. Error bars in b) and c) are 1 s.d.}
    \label{fig5}
\end{figure}

We now demonstrate the complete quantum memory-assisted remote phase sensing protocol. 
Fig.~\ref{fig4}a,b show the experimental setup and circuit. We first generate entanglement between the nuclear qubits at each station using SMPHONE gates, reading out the electron qubit states mid-circuit after the SMPHONE gate application to detect errors. We then reset the electron spin qubits on the Bloch sphere equator:  $\ket{+_{e_L}+_{e_R}}\ket{\Psi^-}_{n_L, n_R}$. After this, the signal light is collected by performing an SMSPG at each station (Fig.~\ref{fig4}b), resulting in :
\begin{equation}
\label{eq:state_after_collection}
    \begin{split}
        [\ket{0_L0_R}\ket{+_{e_L}+_{e_R}} 
        +\sqrt{\mu/2} (\ket{0_L1_R}\ket{+_{e_L}\uparrow_{e_R}}  \\
        +e^{i\phi}\ket{1_L0_R}\ket{\uparrow_{e_L} +_{e_R}})]\ket{\Psi^-}_{n_L, n_R}.
    \end{split}
\end{equation}
We then erase the reflected photonic mode information with feedback $Z_{n_L}^{(1-\mathrm{sign}[i-i'])/2}Z_{n_R}^{(1-\mathrm{sign}[j-j'])/2}$
on the nuclei. To implement non-local signal photon heralding, we apply local 2-qubit gates to entangle the electron and nuclear qubits, and then measure the electron qubits and keep only the outcomes where the two-qubit parity is even ($\ket{\uparrow\uparrow}$ or $\ket{\downarrow\downarrow}$). The vacuum state yields parity outcomes $\ket{\uparrow\downarrow}$ or $\ket{\downarrow\uparrow}$ and thus these measurements can be discarded (see Methods). We note that the non-local heralding is specifically enabled by the entanglement between the stations, without which the signal photon which-path information would be revealed by the heralding step. We measure the non-local heralding probability as a function of $\mu_{\mathrm{sig}}$, which scales with the probability of at least one photon arriving at the stations (Fig.~\ref{fig4}c).

The resulting nuclear state $(\ket{\downarrow\uparrow}-e^{\pm i\phi}\ket{\uparrow\downarrow})/\sqrt{2}$ contains the phase information $\phi$ that can be extracted by evaluating the 2-qubit nuclear $XX$ parity through local measurements (Fig.~\ref{fig4}b,d). 
Figure \ref{fig4}d shows that the nuclear parity amplitude improves with the non-local signal photon heralding, clearly demonstrating the benefit of filtering out the vacuum fluctuation noise. By changing the signal photon number $\mu_{\mathrm{sig}}$ and measuring the nuclear $XX$ parity without non-local heralding, we find the visibility decreases with smaller $\mu_{\mathrm{sig}}$ where vacuum contributions dominate. With non-local heralding, the visibility improves to $0.090(26)$ averaged over all $\mu_{\mathrm{sig}}$ in Fig.~\ref{fig4}e (from $0.031(18)$ without heralding). In an ideal setting, we would expect the signal visibility to remain roughly constant as a function of $\mu_{\mathrm{sig}}$ (with an expected decrease due to multi-photon contamination for large $\mu_{\mathrm{sig}}\gtrsim 1$), but due to errors in the pre-generated Bell state, there is a probability ($\sim 30\%$) to mis-herald photons (equivalent to dark-count ``detection'' of a photon when there was none, red curve in Fig.~\ref{fig4}e), which re-introduces sensitivity to vacuum fluctuations at lower $\mu_{\mathrm{sig}}$.
The visibility enhancement due to vacuum filtering translates to an improved interferometric SNR scaling~\cite{Tsang2011} (Extended Data Fig.~\ref{SIfig_SNR}, see Methods). We note however that mis-heralding events reduce the SNR scaling when the effective mis-heralding probability $\tilde{\varepsilon}_{\mathrm{mh}}$ is larger than the signal $\mu_{\mathrm{sig}}$. Improving the Bell state fidelity is thus central to increasing the range for which optimal SNR can be maintained in the weak signal regime (Extended Data Tab. \ref{SItable:vis_budget}, see Methods).

Finally, we extend the effective baseline by placing spools of fiber between the stations within the entanglement interferometer (Fig.~\ref{fig5}a). The additional length of fiber increases the phase noise of the entanglement interferometer, worsening the locking performance. We maintain a nuclear qubit Bell pair fidelity well above the verifiable entanglement limit with $F=0.63(3)$ for an inter-station fiber length of 1.55~km (3.1~km fiber length inside the interferometer) (Fig.~\ref{fig5}b). Repeating the non-local phase sensing protocol (Fig.~\ref{fig4}b) with a fiber length of 1.55 km, we measure a $\phi$-dependent 2-qubit nuclear parity oscillation visibility of $0.11(4)$ (Fig.~\ref{fig5}c). Since after the entanglement generation all operations are local, these steps are not affected by the increased distance between the stations, although there is of course an increased overhead resulting from the entanglement generation itself.

\section{Outlook}

Our experiments demonstrate entanglement-assisted non-local interferometry by combining key ingredients and techniques, including photon erasure and non-local, non-destructive photon heralding, an enabling step for removing vacuum fluctuations to achieve optimal phase measurement sensitivity~\cite{Tsang2011}. While in our proof-of-concept setting we can extend the measurement baseline to 1.55 km, demonstrating the ability to realize long-baseline quantum memory-assisted interferometry, a number of system improvements will be necessary in order to achieve practical gains over a large baseline. Specifically, 
entanglement rates should be increased using quantum repeaters~\cite{Azuma2023} and entanglement multiplexing~\cite{Ruskuc2025, Collins2007}. The number of qubits at each station can be increased with additional devices per station and by incorporating additional quantum memories with $^{13}$C nuclear spin control \cite{Nguyen2019PRL, Bradley2019}. This would allow for efficient storage of incoming photons enabling the logarithmic compression method described in \cite{emil_telescopes}, and the extraction of both temporal \cite{Vaughan2013} and polarization \cite{Hough2006} information. Furthermore, utilizing spin-photon phase gates instead of amplitude gates used here opens pathways to higher efficiency deterministic operations (see Supplementary Information). The Purcell-enhanced $\sim 1$ GHz linewidth of the SiV constrains the spectral window, but this can be substantially extended by increasing the number of devices per station with efficient and scalable fiber packaging~\cite{Zeng2023}, combined with wavelength division multiplexing~\cite{Ciurana2014}, and strain-induced SiV optical frequency tuning~\cite{Machielse2019}.

Our experiments establish a novel approach for advancing quantum-enhanced optical imaging by demonstrating coherent storage and manipulation of weak optical signals using quantum devices. Encoding such signals into qubit memories and coupling them to modest-scale quantum processing units can further allow for application of advanced quantum algorithms to extract information beyond the reach of direct detection and classical post-processing. For example, by extending the present approach to multiple detectors quantum algorithms can be utilized to overcome tomographic constraints and shot-noise accumulation in classical techniques, leading to fundamental improvements of SNR scaling with system dimensionality~\cite{Huang2022,exoplanets}. For instance such systems can be used to improve the performance in demanding imaging tasks such as exoplanet detection~\cite{exoplanets}. 
Therefore, our experiments open the door for realizing quantum-enhanced imaging in the weak-signal regime, with potentially transformative applications ranging from curved spacetime proper-time interferometry~\cite{Borregaard2025} and deep-space optical communication~\cite{Cui2025, Smith2025} to more general weak-signal imaging tasks \cite{Casacio2021, Thekkadath2023, Allen2025}.

\bibliography{main_arxiv}
\pagebreak
\onecolumngrid
\pagebreak
\twocolumngrid

\section{Methods}

\subsection{Experimental setup}
The experiment encompasses two labs, each containing one station with an SiV inside a dilution refrigerator (BlueFors BF-LD250) at $\sim 100$~mK and connected as shown in Extended Data Fig.~\ref{SIfig_setup}. Light signals are prepared in the laser setup, including for entanglement and erasure interferometer locking (NewFocus TLB-6700 Velocity), right station SiV readout, entanglement qubit generation, signal light generation, and erasure LO generation (MSquared SolsTiS Ti:Sapphire), left station SiV readout and filter cavity locking (Toptica DLPro) and SiV de-ionization (Thorlabs Green diode LP520-SF15). All free space and in-fiber acousto-optic modulators (AOM) are driven with 215~MHz. The entanglement qubit, signal light, and erasure LO pulses are shaped by modulating the RF signal sent to in-fiber AOMs. We bridge the frequency difference between the SiVs at each station of $\Delta f_{L-R}\sim 10$~GHz (Extended Data Fig.~\ref{SIfig_SiV_properties}) by generating sidebands with electro-optic modulators (EOM) driven with a radio-frequency signal at $\Delta f_{L-R}$ inside the entanglement interferometer and filtering the light with a Fabry-Perot cavity \cite{two_fridge}. Free space AOMs at the left and right stations act both as a switch between entanglement path and signal-erasing path, as well as frequency shifters for entanglement interferometer phase locking. Photon counts for erasure are detected with pairs of superconducting nanowire single photon detectors (SNSPD) (Photon Spot) at each station, and entanglement photon heralding clicks are detected with a single-photon avalanche photodiode (APD). We note that the erasure SNSPDs are not instantaneously photon-number-resolving but can effectively resolve photon number when the detector deadtime is much lower than the photon length. All counts are logged with a time tagger (Swabian Instruments Time Tagger Ultra), and two Zurich Instrument HDAWG 2.4~GSa/s arbitrary waveform generators are used for sequence logic, control of the AOMs and EOMs, as well as MW and RF pulse generation for SiV control.

\subsection{Signal-to-noise ratio and Fisher information}
\label{SI_fisher_info}

The SNR can be evaluated through the Fisher information $\mathcal{F}_I$ of the measurement, which is equivalent to $(\mathrm{SNR})^2$ and bounds the $\phi$ estimation variance as $\mathrm{Var}(\phi_{\mathrm{est}}) \geq 1/\mathcal{F}_I$ \cite{Tsang2011, Vittorio2011}:

\begin{equation}
\label{eq:fisher_info}
    \mathcal{F}_I = \sum_y\frac{1}{P(y|\phi)}\left ( \frac{\delta P (y|\phi)}{\delta \phi} \right )^2,
\end{equation}
where for our experiment $P(y|\phi)$ is the probability of obtaining a nuclear 2-qubit parity measurement outcome $y$ for a given $\phi$. The probabilities $P(y | \phi)$ are:

\begin{equation}
\label{eq:outcomes}
\begin{split}
    P(\mathrm{discarded}|\phi) = 1-p_{\mathrm{succ}} \\
    P(\pm\,\mathrm{parity}|\phi) = p_{\mathrm{succ}} (1\pm V\mathrm{cos}(\phi))/2,
\end{split}
\end{equation}
where $V$ is the visibility of the measurement and $p_{\mathrm{succ}}$ is the probability to herald a photon (including the photon presence probability itself). Using Eq. \ref{eq:fisher_info} and \ref{eq:outcomes} for small visibility $V^2\ll 1$, we get $\mathcal{F}_I \propto p_{\mathrm{succ}}V^2$. 

For our implemented protocol, $p_{\mathrm{succ}}=\eta_{\mathrm{erasure}}\eta_{\mathrm{herald}}P(n_{\mathrm{photon}}\geq 1)$, where $\eta_{\mathrm{erasure(herald)}}$ are constant factors given by the erasure (signal photon heralding) efficiency. The signal photon heralding efficiency is limited to 50\% by the use of amplitude-based SMSPG, but can be increased to 100\% by using phase-based gates instead (see Supplementary Information). The sequence heralds whether there was at least one signal photon but does not distinguish between single and multi-photon events. Since the protocol fails when more than one photon is collected, the visibility is given by $V=\overline{V} P(n_{\mathrm{photon}}=1| n_{\mathrm{photon}}\geq 1)$ (Fig.~\ref{fig4}e, dashed red curve), where $\overline{V}$ is the constant overhead factor due to fidelity reduction from imperfect photon erasure, gate errors, and initial qubit state fidelities. For a light signal with average photon number $\mu_{\mathrm{sig}}$ arriving at the stations, the Fisher information is 
\begin{equation}
    \mathcal{F}_I =  \eta_{\mathrm{erasure}}\eta_{\mathrm{herald}}\overline{V}^2\frac{\mu_{\mathrm{sig}}^2e^{-2\mu_{\mathrm{sig}}}}{1-e^{-\mu_{\mathrm{sig}}}},
\end{equation}
which reduces to $\mathcal{F}_I \propto \mu_{\mathrm{sig}}$ for small signal $\mu_{\mathrm{sig}}\ll 1$.

Without signal photon heralding, $p_{\mathrm{succ}}=\eta_{\mathrm{erasure}}$ and $V=\eta_{\mathrm{herald}}\overline{V} P(n_{\mathrm{photon}}=1)$ (Fig.~\ref{fig4}e, solid blue curve), so that the resulting Fisher information is:

\begin{equation}
    \mathcal{F}_I =  \eta_{\mathrm{erasure}}(\eta_{\mathrm{herald}}\overline{V})^2\mu_{\mathrm{sig}}^2e^{-2\mu_{\mathrm{sig}}},
\end{equation}
which reduces to $\mathcal{F}_I \propto \mu_{\mathrm{sig}}^2$ for $\mu_{\mathrm{sig}}\ll 1$. This precisely shows that the key feature that enables SNR scaling enhancement is the non-destructive non-local signal photon heralding. This step, enabled by pre-generated entanglement, is what gives the remote phase sensing protocol its non-local character. 

On the other hand, when using non-local signal photon heralding, mis-heralding events (with probability $\varepsilon_{\mathrm{mh}}$) corrupt the signal, modifying $p_{\mathrm{succ}}$ to $\eta_{\mathrm{erasure}}P(\mathrm{herald}\,\cup \,\mathrm{mis-herald})$ and $V$ to $\overline{V} P(n_{\mathrm{photon}}=1| (\mathrm{herald}\,\cup \,\mathrm{mis-herald}))$ (Fig.~\ref{fig4}e, solid red curve). This results in:

\begin{equation}
    \mathcal{F}_I =  \eta_{\mathrm{erasure}}\eta_{\mathrm{herald}}\overline{V}^2\frac{\mu_{\mathrm{sig}}^2e^{-2\mu_{\mathrm{sig}}}}{1-e^{-\mu_{\mathrm{sig}}}(1-\varepsilon_{\mathrm{mh}}/\eta_{\mathrm{herald}})},
\end{equation}

which scales as $\mathcal{F}_I \propto \mu_{\mathrm{sig}}$ for $\mu_{\mathrm{sig}}\gtrsim \varepsilon_{\mathrm{mh}}/\eta_{\mathrm{herald}}$ but curves down to $\mathcal{F}_I \propto \mu_{\mathrm{sig}}^2$ for $\mu_{\mathrm{sig}}\lesssim \varepsilon_{\mathrm{mh}}/\eta_{\mathrm{herald}}$ (Extended Data Fig. \ref{SIfig_SNR}, see Supplementary Information). The visibility improvement from signal photon heralding can be seen both in Fig.~\ref{fig4}d and Extended Data Fig.~\ref{SIfig_timebin_heralding}.

\subsection{SMPHONE gate error detection}

Similarly to the PHONE gate \cite{si29_node, two_fridge}, the SMPHONE gate entangles a photonic qubit with the nuclear spin -- but in the Fock basis instead of the time-bin basis -- mediated by the electron spin. Starting the nucleus and the photon in superposition states $(\ket{\downarrow}+\ket{\uparrow})_n/\sqrt{2}$ and $(\ket{0}+\ket{1})_{\mathrm{phot}}/\sqrt{2}$ and the electron in the $\ket{\uparrow}_e$ state, we implement the SMPHONE gate (Extended Data Fig.~\ref{SIfig_SMPHONE}a):

\begin{multline}(\ket{0}+\ket{1})_{\mathrm{phot}}\ket{\uparrow}_e(\ket{\downarrow}+\ket{\uparrow})_n \\
\longrightarrow \ket{\uparrow}_e(\ket{0}_{\mathrm{phot}}\ket{+}_n +\ket{1}_{\mathrm{phot}}\ket{\downarrow}_n/\sqrt{2}).
\end{multline}
Here the nucleus is entangled with the photon and the electron is always in the $\ket{\uparrow}$ state, unless a MW error occurred during the SMPHONE gate operation. Therefore, by measuring the electron state we can detect MW errors and post-select on $\ket{\uparrow}$ results to boost the nucleus entanglement fidelity (Extended Data Fig.~\ref{SIfig_SMPHONE}b). Since measuring the electron in the $\ket{\uparrow}$ state (as opposed to the $\ket{\downarrow}$ state) does not cause decoherence of the $^{29}$Si nucleus state \cite{si29_node}, we can perform error detection mid-circuit, as in Fig.~\ref{fig4}b.

\subsection{Entanglement interferometer phase}

The entanglement interferometer phase $\delta\varphi_e$ stability (Extended Data Fig.~\ref{SIfig_entanglement_phase_stab}c) is limited by noise from the fiber link between the two labs where the stations are located and vibrations from the pulse tube motor-head of the dilution refrigerators. We reduce phase noise introduced in the fiber link by packaging the fiber in a rubber tube filled with sand for vibration damping (Extended Data Fig.~\ref{SIfig_entanglement_phase_stab}a). We limit the phase noise introduced by the pulse tube motor-head by clamping the motor-head between aluminum plates padded with foam. To reduce vibrations guided to the dilution refrigerator through the flexline connecting to the pulse tube head, we clamp the flexline in bags of sand that further damp vibrations (Extended Data Fig.~\ref{SIfig_entanglement_phase_stab}b). With this passive stabilization the interferometer phase autocorrelation time increases from $\sim 4$ ms to $\sim 500$ ms (Extended Data Fig.~\ref{SIfig_entanglement_phase_stab}c). When we add the two spools of 1.5 km for the long-baseline operation (Fig.~\ref{fig5}), the auto-correlation time of the entanglement interferometer decreases again (Extended Data Fig.~\ref{SIfig_entanglement_phase_stab}c, inset). 

We then lock the interferometer phase by alternating phase probing with SiV readout every 50 $\mu$s. A field-programmable gate array (FPGA) integrates the phase probing light for 1 ms and locks the interferometer phase by adjusting the drive frequency of Acousto-Optic Modulators (AOM) in each arm, resulting in a locked optical interference visibility of $\sim 0.93$ (Extended Data Fig.~\ref{SIfig_entanglement_phase_stab}d).

\subsection{Quantum memory-assisted interferometry details}
After generating entanglement and collecting the signal in the non-local phase sensing protocol, with the resulting state in Eq. \ref{eq:state_after_collection}, we erase the photonic spatial mode:

\begin{multline}
    [\ket{+_{e_L}+_{e_R}} \\
    +\sqrt{\mu_{\mathrm{sig}}/2} (\ket{+_{e_L}\uparrow_{e_R}} +e^{i\phi}\ket{\uparrow_{e_L} +_{e_R}})]\ket{\Psi^-}_{n_L, n_R}.
\end{multline}
Then, with local C$_n$NOT$_e$ and $\pi/2$ pulses at each node, we transform the state to:

\begin{multline}
    (\ket{\downarrow_{e_L}\uparrow_{e_R}}\ket{\downarrow_{n_L}\uparrow_{n_R}}-\ket{\uparrow_{e_L}\downarrow_{e_R}}\ket{\uparrow_{n_L}\downarrow_{n_R}})/\sqrt{2} \\
    +\sqrt{\mu}/2 [ \ket{\downarrow_{e_L}\downarrow_{e_R}}(\ket{\downarrow_{n_L}\uparrow_{n_R}}-e^{i\phi}\ket{\uparrow_{n_L}\downarrow_{n_R}}) \\
    + \ket{\uparrow_{e_L}\uparrow_{e_R}}(e^{i\phi}\ket{\downarrow_{n_L}\uparrow_{n_R}}-\ket{\uparrow_{n_L}\downarrow_{n_R}}) \\
    +\ket{\downarrow_{e_L}\uparrow_{e_R}}(1+e^{i\phi})\ket{\downarrow_{n_L}\uparrow_{n_R}}+\ket{\uparrow_{e_L}\downarrow_{e_R}}(1-e^{i\phi})\ket{\uparrow_{n_L}\downarrow_{n_R}}],
\end{multline}

so the electron two-qubit parity is even ($\ket{\uparrow\uparrow}$ or $\ket{\downarrow\downarrow}$) only if a signal photon was present, and the probability of measuring these states scales with the probability of at least one signal photon arriving (Fig.~\ref{fig4}c). We note that the mis-heralding probability is higher for heralding on the $\ket{\downarrow_e\downarrow_e}$ than the $\ket{\uparrow_e\uparrow_e}$ electron state due to experimental errors accumulating coherently in the $\ket{\downarrow_e\downarrow_e}$ state. The nuclear 2-qubit parity oscillation curves with and without non-local signal photon heralding in function of signal phase separated by signal strength are shown in Extended Data Fig.~\ref{SIfig_additional_telescope_data}. These curves are combined to plot the curve in Fig.~\ref{fig4}e.

The phases of the LO pulses used in the photon erasure step at the left and right stations are also imprinted onto the nuclear state, so that the relevant phase is $\Delta\Phi_{L-R} = \delta\phi_L - \delta\phi_R$ with $\delta\phi_{L(R)}$ the differential phase between the signal and the LO pulses at the left (right) station. $\Delta\Phi_{L-R}$ reduces to simply $\phi$ when the phases of the LO pulses are locked to one another. However, it is enough to simply know the phases through calibration measurements which we perform every 4 experimental trials (See Supplementary Information).

\subsection{Coherent state thermalization}
For local phases $\delta\varphi_{L(R)}$ at the left (right) station, and differential phase $\phi=\delta\varphi_L-\delta\varphi_R$, the photonic state density matrix in the $\{\ket{00},\ket{01},\ket{10}\}$ bases for weak signals $\mu\ll 1$ is:

\begin{equation*}
\rho_{\mathrm{sig}} \approx
\begin{pmatrix}
1 & e^{-i\delta\varphi_R}\sqrt{\mu/2} & e^{-i\delta\varphi_L}\sqrt{\mu/2} \\
e^{i\delta\varphi_R}\sqrt{\mu/2} & \mu/2 & e^{-i\phi} \mu/2  \\
e^{i\delta\varphi_L}\sqrt{\mu/2} & e^{i\phi}\mu/2 & \mu/2
\end{pmatrix}
.
\label{eq:rho_coh_mu}
\end{equation*}

We fix $\phi=\delta\varphi_L-\delta\varphi_R$, and sample over a uniform random distribution of local phases $\delta\varphi_L$ and $\delta\varphi_R$ (Extended Data Fig.~\ref{SIfig_phase_dist}), so that the density matrix becomes

\begin{equation*}
\tilde{\rho}_{\mathrm{sig}} \approx 
\begin{pmatrix}
1 & 0 & 0 \\
0 & \mu/2 & e^{-i\phi} \mu/2  \\
0 & e^{i\phi}\mu/2 & \mu/2
\end{pmatrix},
\label{eq:rho_coh_mu_therm}
\end{equation*}
which is equivalent to a weak thermal state density matrix (with mean photon number $\mu$) \cite{Tsang2011}.

\subsection{Acknowledgements}
We thank Mihir Bhaskar, David Levonian, Denis Sukachev, Madison Sutula, Elana Urbach, and Nicholas Mondrik for useful discussions and experimental help, and Chawina De-Eknamkul for support with the tapered-fiber-optical interface. This work was supported by the AWS Center for Quantum Networking, the National Science Foundation (Grant No. PHY-2012023), NSF Center for Ultracold Atoms, the NSF Engineering Research Center for Quantum Networks (Grant No. EEC-1941583), CQN (EEC-1941583), and NSF QuSeC-TAQS OMA2326787. Devices were fabricated at the Harvard Center for Nanoscale Systems, NSF award no. 2025158. G.B. acknowledges support from the MIT Patrons of Physics Fellows Society. Y.Q.H acknowledges support from the A*STAR National Science Scholarship.
\subsection{Author contributions}
P.-J.S., Y.-C.W., M.S., and A.S. planned the experiment. U.Y., B.M., and D.R. fabricated the devices. P.-J.S., Y.-C.W., M.S., Y.Q.H., F.A.A., E.K., and A.S. built the setup and P.-J.S., Y.-C.W., and M.S. performed the experiment. P.-J.S., Y.-C.W., M.S., Y.Q.H., F.A.A., E.K., S.G., and G.B. analyzed the data and interpreted the results. All work was supervised by H.P., M.L., J.B., A.S., and M.D.L. All authors discussed the results and contributed to the manuscript. P.-J.S., Y.-C.W., and M.S. contributed equally to this work.
\subsection{Supplementary information}
Supplementary Information is available for this paper.
\subsection{Data availability}
All data related to the current study are available from the corresponding author upon request.
\subsection{Code availability}
All analysis code related to the current study are available from the corresponding author upon request.
\subsection{Competing interests}
M.D.L. is a co-founder, shareholder, and chief scientist of QuEra Computing. B.M. is director photonics technologies of IonQ. D.R. is a senior research science manager of IonQ. J.B. is a senior staff research scientist of IonQ.
\subsection{Corresponding author}
All correspondence should be addressed to M.D.L.

\setcounter{figure}{0}
\renewcommand{\figurename}{Extended Data Fig.}
\renewcommand{\tablename}{Extended Data Table}

\pagebreak
\onecolumngrid
\pagebreak
\begin{figure}[h]
    \centering
    \includegraphics[width=0.8\textwidth]{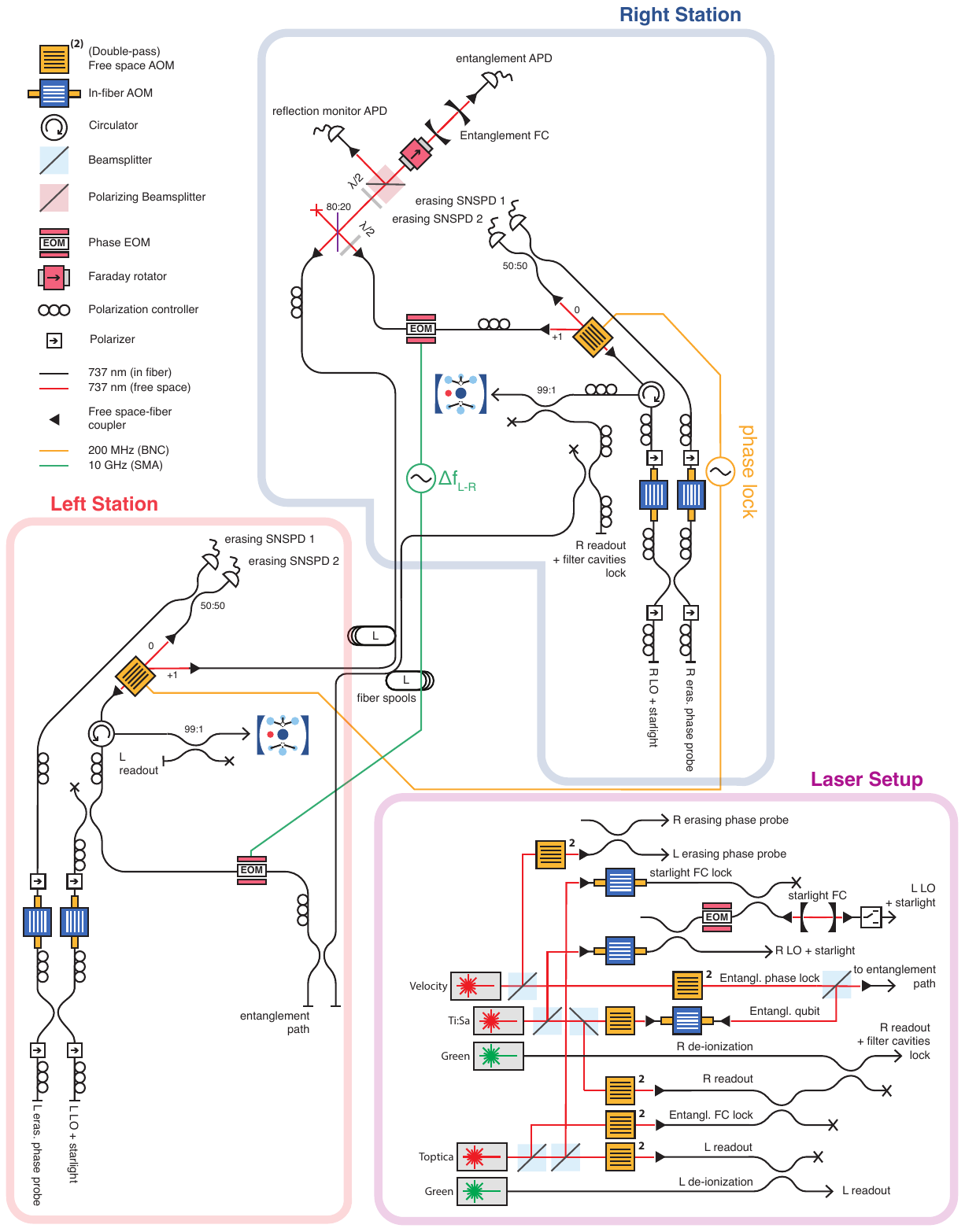}
    \caption{\textbf{Experimental setup.} The Toptica laser is set to the left station SiV optical frequency $f_L$, the Ti:Sa laser to the right station SiV optical frequency $f_R$ and the Velocity laser to $f_R + 5\cdot\mathrm{FSR}$, where $\mathrm{FSR}$ is the free spectral range of the entanglement filter cavity of $\sim 46$ GHz. APD and SNSPD stand for avalanche photodiode and superconducting nanowire single-photon detector, and FC is filter cavity.}
    \label{SIfig_setup}
\end{figure}

\begin{figure}[h]
    \centering
    \includegraphics[width=0.35\textwidth]{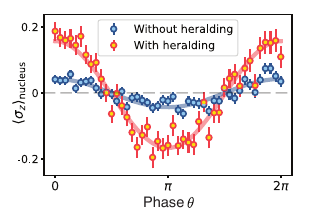}
    \caption{\textbf{Impact of photon heralding on single node temporal mode phase sensing.} Nuclear state measurement result after the sequence shown in Fig.~\ref{fig3}a with (red) and without (blue) signal photon heralding.}
    \label{SIfig_timebin_heralding}
\end{figure}

\begin{figure}[h]
    \centering
    \includegraphics[width=0.35\textwidth]{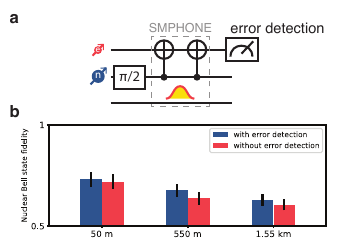}
    \caption{\textbf{SMPHONE gate operation.} a)~SMPHONE gate sequence, after which the electron state can be measured to detect errors. b)~Nuclear Bell state fidelity for different baseline lengths with and without error detection.}
    \label{SIfig_SMPHONE}
\end{figure}

\begin{figure}[h]
    \centering
    \includegraphics[width=0.4\textwidth]{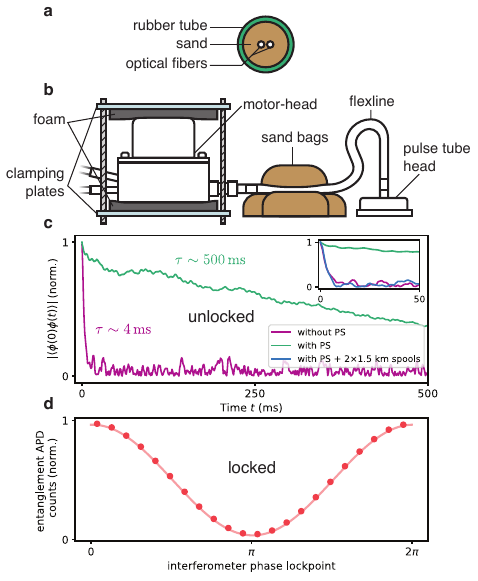}
    \caption{\textbf{Entanglement interferometer phase passive stabilization and active locking.} a)~Fiber packaging used for the inter-station labs fiber link. b)~Pulse tube motor-head noise isolation. c)~Entanglement interferometer accumulated phase auto-correlation function with (green) and without (purple) passive phase stabilization. Inset: zoomed-in auto-correlation curves, including for the interferometer with 2 spools of 1.5 km each. d)~Entanglement APD counts as a function of locked interferometer phase lock setpoint.}
    \label{SIfig_entanglement_phase_stab}
\end{figure}

\begin{figure}[h]
    \centering
    \includegraphics[width=0.35\textwidth]{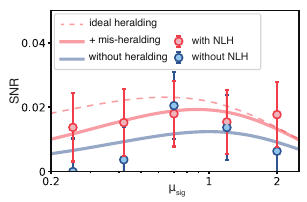}
    \caption{\textbf{Signal to noise ratio (SNR) of the remote phase measurement in funtion of $\mu_{\mathrm{sig}}$} with (in red) and without (in blue) non-local heralding. The SNR is calculated by taking the square root of the Fisher information $\sqrt{\mathcal{F}_I}$. The solid blue (dashed red) curve corresponds to the SNR scaling without (with ideal) heralding, and the solid red curve corresponds to the scaling with imperfect heralding due to mis-heralding.}
    \label{SIfig_SNR}
\end{figure}

\begin{figure}[h]
    \centering
    \includegraphics[width=\textwidth]{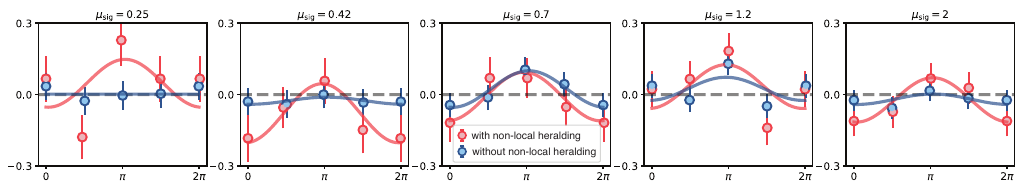}
    \caption{\textbf{Additional data for implementation of quantum memory-assisted interferometry.} For each average photon number $\mu_{\mathrm{sig}}$, the plot shows the nuclear 2-qubit parity $\langle XX\rangle$ in function of signal phase $\Delta\Phi_{L-R}$ with (in red) and without (in blue) non-local heralding.}
    \label{SIfig_additional_telescope_data}
\end{figure}

\begin{figure}[h]
    \centering
    \includegraphics[width=0.6\textwidth]{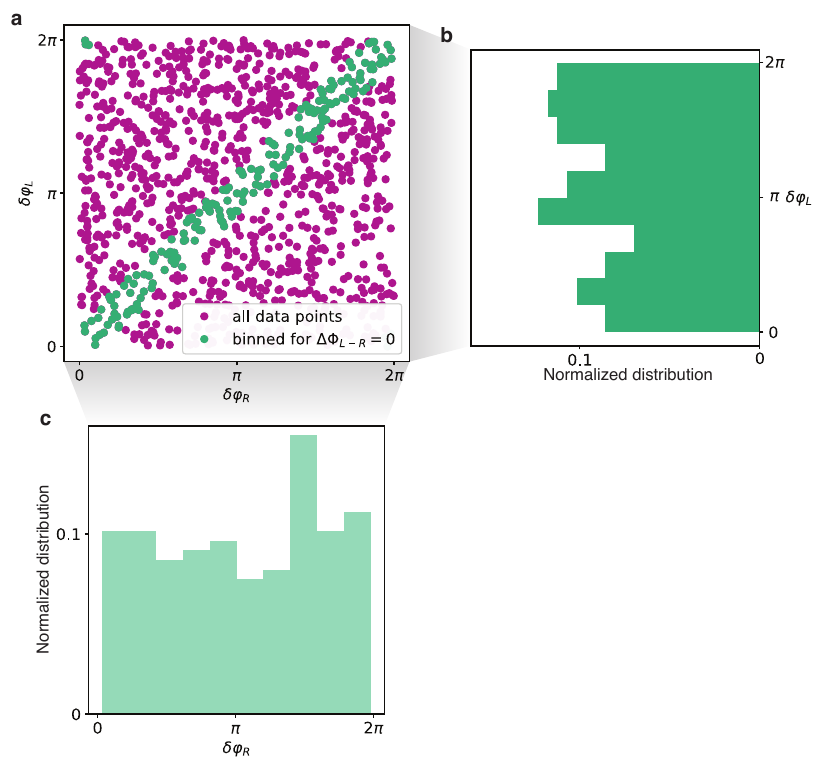}
    \caption{\textbf{Uniform distribution of local phases.} a)~Scatter plot of all local phases in purple, and for the local phases for the binning of differential phases $\Delta\Phi_{L-R}=0$. b) and c)~are histograms of the local phases for the left ($\delta\varphi_L$) and right ($\delta\varphi_R$) for the binning of differential phases $\Delta\Phi_{L-R}=0$, respectively. }
    \label{SIfig_phase_dist}
\end{figure}

\begin{figure}[h]
    \centering
    \includegraphics[width=0.8\textwidth]{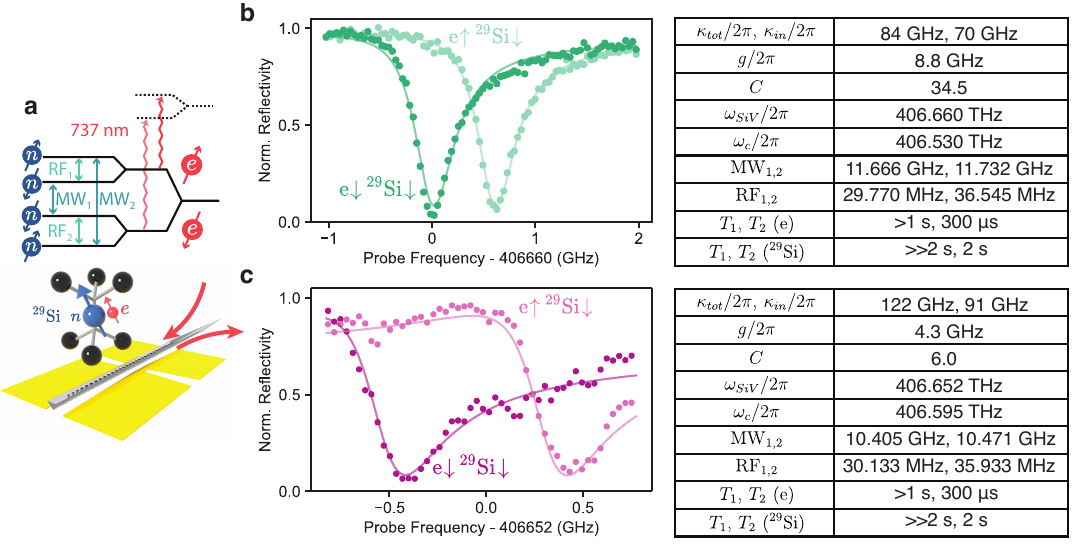}
    \caption{\textbf{Cavity-QED parameters and SiV properties} a)~Energy level structure of the SiV b)~Left station and c)~Right station. Left: Measured normalized reflectivity as a function of optical frequency. 
    Right: Extracted cavity QED parameters and transitions: total cavity decay constant $\kappa_{\text{tot}}$, cavity input decay constant $\kappa_{\text{in}}$, SiV-cavity coupling strength $g$, cavity resonance frequency~$\omega_c$, SiV resonance frequency $\omega_{\text{SiV}}$ (for $\ket{\downarrow}$), cooperativity $C$, microwave transitions MW$_1$, MW$_2$, and radio-frequency transitions RF$_1$, and RF$_2$.}
    \label{SIfig_SiV_properties}
\end{figure}

\begin{table}[h]
	\begin{center}
		\begin{tabular}{c c c}
              & Time-bin phase sensing & Spatial phase sensing\\
             \hline
              Initial state preparation & 0.9 ($V_Z$)& 0.64 ($V_{XX}$)\\
              Mis-heralding ($1-p_{\mathrm{mh}}$) & 0.45 & $0.32 (\mu_{\mathrm{sig}}=0.25)-0.91 (\mu_{\mathrm{sig}}=2)$ \\
              Erasure (from loss) & 0.5 ($\alpha_{\mathrm{LO}}=0.45$) & 0.4 ($\alpha_{\mathrm{LO}}=0.55$) \\
              Multi-photon error & 0.95 ($\mu_{\mathrm{sig}}=1$) & $0.88 (\mu_{\mathrm{sig}}=0.25)-0.31 (\mu_{\mathrm{sig}}=2)$ \\
              \hline
              Expected visibility & 19\% & 7.2\% \\

              Measured visibility & 16(1)\% & 9.0(2.6)\%\\
              \hline
        \end{tabular}
	\end{center}
	\caption{\textbf{Visibility reduction from different sources.} Time-bin sensing corresponds to the experiment in Fig.~\ref{fig3}, and Spatial (non-local) sensing corresponds to the experiment in Fig.~\ref{fig4}. $V_Z$ is the $Z$-basis visibility of the initial nucleus $\ket{+}$ state, and $V_{XX}$ is the $XX$-basis visibility of the initial nuclei $\ket{\Psi^-}$ state.}
	\label{SItable:vis_budget}
\end{table}

\pagebreak
\twocolumngrid
\pagebreak
\onecolumngrid
\pagebreak

\clearpage



\setcounter{equation}{0}
\setcounter{figure}{0}
\setcounter{table}{0}
\setcounter{page}{1}
\setcounter{section}{0}
\makeatletter

\renewcommand{\figurename}{Fig.}
\renewcommand{\tablename}{Table}

\renewcommand{\theequation}{S\arabic{equation}}
\renewcommand{\thefigure}{S\arabic{figure}}
\renewcommand{\thetable}{S\arabic{table}}

\pagebreak
\begin{center}
\large Supplementary Information for\\\textbf{Entanglement Assisted Non-local Optical Interferometry in a Quantum Network}
\end{center}

\section{Comparison of interferometric phase measurement schemes}
\begin{figure*}[ht]
    \centering
    \includegraphics[width=0.85\textwidth]{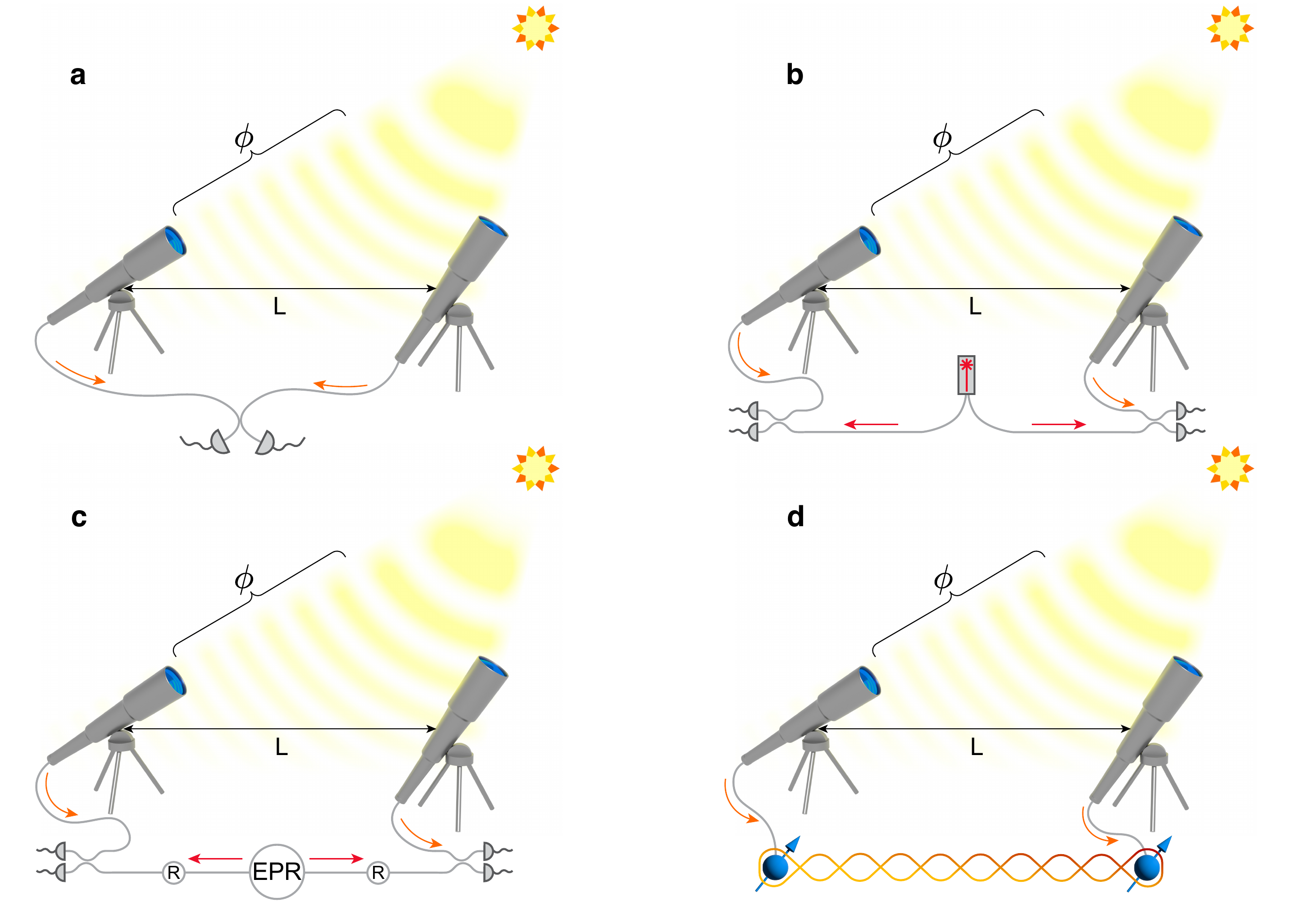}
    \caption{\textbf{Four primary schemes for long-baseline optical interferometry to measure signal light phase $\phi$ over a distance $L$.} 
    a) Direct detection: signal light is combined on a common beamsplitter. 
    b) Local homodyne or heterodyne detection: signal light interferes with a local oscillator. 
    c) GJC scheme: employs path-entangled single-photon source (EPR pairs) and quantum repeaters (R). 
    d) KBGL scheme: utilizes quantum memory nodes and shared Bell pairs. 
    All schemes except (b) are non-local and exhibit linear Fisher information scaling. Scheme (b) is local and scales quadratically, making it suboptimal for weak signals.}
    \label{SIfig_all_schemes}
\end{figure*}

We consider four principal schemes for long-baseline interferometry, illustrated in Fig.~\ref{SIfig_all_schemes}:

\begin{enumerate}[label=(\alph*)]
    \item \textbf{Direct detection} — combines signal light on a common beamsplitter for direct interference.
    
    \item \textbf{Local homodyne and heterodyne detection} — measures interference between signal light and a distributed local oscillator at each node.
    
    \item \textbf{Gottesman–Jennewein–Croke (GJC) scheme} — employs path-entangled single-photon source (e.g., EPR pairs) and quantum repeaters to distribute entanglement non-locally \cite{gottesman2012}.
    
    \item \textbf{Khabiboulline–Borregaard–de Greve–Lukin (KBGL) scheme} — utilizes quantum memory nodes and shared Bell pairs to enable non-local phase-sensitive measurements \cite{emil_telescopes}, experimentally shown in this work.
\end{enumerate}

All schemes except for (b) are considered non-local and exhibit Fisher information that scales linearly with the signal strength. 
Scheme (b), being local, has Fisher information that scales quadratically—rendering it less favorable for detecting weak signals. 
Intuitively, this is because non-local measurements (a, c, d) preserve the coherence of the signal photon's wavefunction, 
allowing interference and enabling one photon to yield one data point. In contrast, local measurements (b) collapse the wavefunction, 
requiring separate phase estimation at each node and thus two photons per data point to obtain the phase difference~$\phi$.

Direct detection on the other hand suffers from exponential transmission loss with the node separation~$L$. 
Local homodyne and heterodyne detection avoid this loss but are hampered by their quadratic Fisher information scaling. 
This motivates the exploration of other non-local schemes enabled by quantum networks: GJC and KBGL, which we now describe and compare.

\begin{figure*}[ht]
    \centering
    \includegraphics[width=1\textwidth]{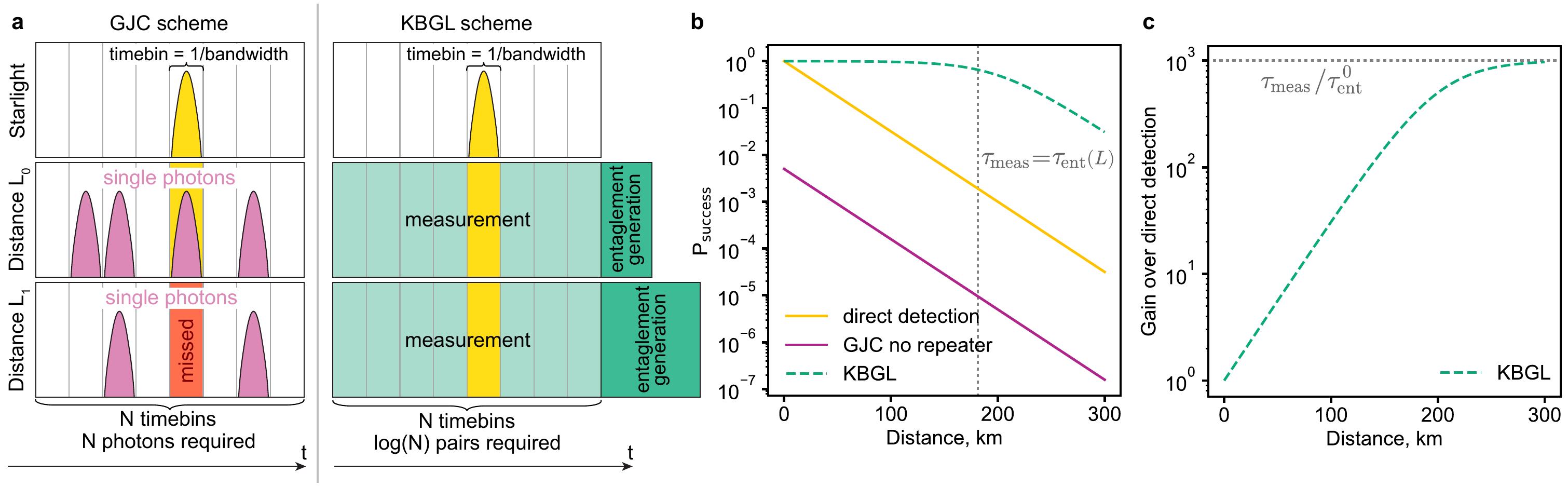}
    \caption{\textbf{Performance of quantum-assisted schemes as a function of node separation.} 
    a)~(left)~GJC scheme: each timebin corresponding to a signal photon must be covered by a single photon from the source, which undergoes transmission losses that increase with distance. 
    (right)~KBGL protocol: a fixed measurement window (capturing, on average, one signal photon) is divided into $N$ timebins, requiring only $\log_2(N)$ memory qubits and the same number of Bell pairs. The duration of the entanglement window increases with distance. 
    b)~Success probability as a function of distance for direct detection, GJC with no repeater (and with current state-of-the-art heralded single-photon source), and KBGL (with a kink when $\tau_{\text{meas}}=\tau_{\text{ent}}(L)$). 
    c)~Gain of KBGL over direct detection. 
    Simulation parameters: $\Delta f = 1$~MHz signal light bandwidth, 1~s measurement window, 20~SiVs per node, and a 1~kHz zero-distance entanglement rate for optimal and realistic KBGL performance, saturating at $\tau_{\text{meas}}/\tau_{\text{ent}}^0$. Standard telecom fiber 0.3~dB/km attenuation is assumed for (b-c).}
    \label{SIfig_schemes_comparison_kbgl}
\end{figure*}

When a signal photon with bandwidth~$\Delta f$ and coherence time~$\tau_{\text{star}} = 1/\Delta f$ arrives at the detector, 
the telescope's measurement window can be divided into time bins of duration~$\tau_{\text{star}}$ (Fig.~\ref{SIfig_schemes_comparison_kbgl}a). 
The full measurement window should be short enough to ensure a low probability of multi-photon events. 
In the GJC scheme, effective quantum interference requires that each time bin consumes a photon, 
necessitating~$N$ single photons for~$N$ timebins from the path-entangled single-photon (EPR pairs) source. However, due to transmission loss, the success probability per signal photon~$P_{\text{success}}$ drops exponentially. 

In the absence of a quantum repeater, the GJC scheme can at best match direct detection performance, assuming an ideal deterministic single-photon source providing one photon per each time bin.
To overcome exponential losses, quantum repeaters are necessary, and a high zero-distance photon rate is still required. 
With current state-of-the-art heralded single-photon sources and no quantum repeater the GJC success rate per signal photon is roughly two orders of magnitude lower than that of direct detection (Fig.~\ref{SIfig_schemes_comparison_kbgl}b).

In contrast, the KBGL scheme does not require repeaters or high Bell pair generation rates to overcome exponential losses with distance. 
In this scheme measurement time bins are stored in quantum memories (Fig.~\ref{SIfig_schemes_comparison_kbgl}a). 
With binary encoding, only $\log_2(N)$ memory qubits are needed to store $N$ timebins, significantly reducing the overhead and enabling longer collection windows, 
limited by memory coherence time. However, the KBGL schemes also requires time for entanglement generation and data processing (non-local parity measurements). 
When the measurement window is larger than the entanglement and processing time, exponential loss is circumvented, and $P_{\text{success}}$ does not depend on distance -- an improvement over direct detection (Fig.~\ref{SIfig_schemes_comparison_kbgl}b).

\begin{figure*}[ht]
    \centering
    \includegraphics[width=1\textwidth]{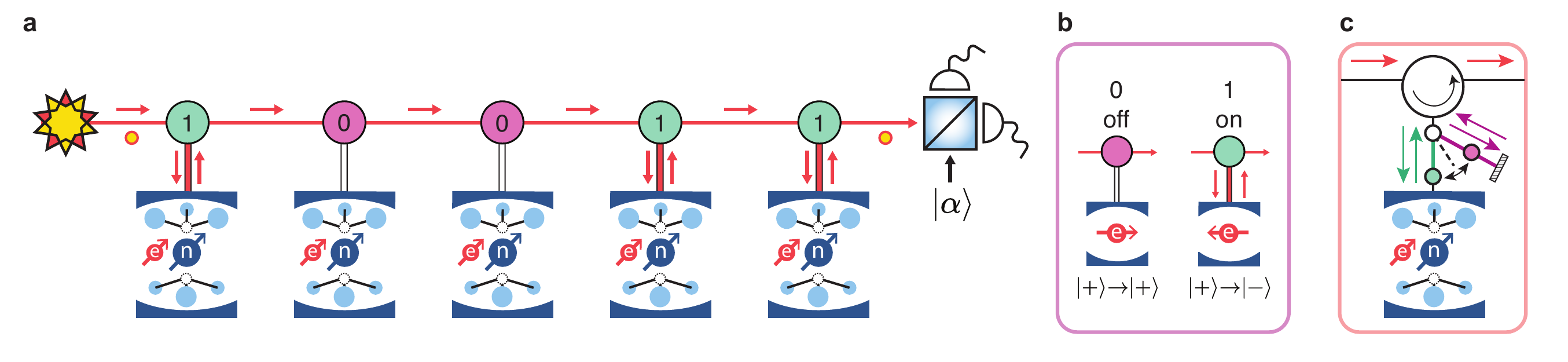}
    \caption{\textbf{Binary loading of a signal photon into quantum memory \cite{zheng2022entanglement}}. a) $N$ timebins require $\log_2(N)$ quantum memories and $\log_2(N)$ switches (in this picture 5 qubits can store $2^5=32$ timebins, and loading of 19$^{\text{th}}$ timebin (10011$_2$) is shown). Photon erasure with coherent state is then performed. b) Green (purple) represents switch in \textit{on} (\textit{off}) state, where the SiV interacts with a photon and thus acquires $\pi$ phase when the switch is in the \textit{on} state. c) Exemplary design of the switch is shown to the right, involving a simple binary optical switch, a retroreflector and a circulator.}
    \label{SIfig_binary_loading}
\end{figure*}

Assuming a signal photon rate~$r_{\text{star}}\approx1/\tau_{\text{meas}}$ and transmission loss scaling as~$e^{-L/(2L_0)}$ (with the beamsplitter midway), 
the direct detection success probability per signal photon is
\[
P_{\text{success}}^{\text{direct}} = e^{-L/(2L_0)}.
\]

For KBGL with a fixed measurement window~$\tau_{\text{meas}}$ and an entanglement duration $\tau_{\text{ent}}(L) = \tau_{\text{ent}}^0 e^{L/(2L_0)}$, 
the success probability per signal photon is
\[
P_{\text{success}}^{\text{KBGL}} = \frac{\tau_{\text{meas}}}{\tau_{\text{meas}} + \tau_{\text{ent}}^0 e^{L/(2L_0)}}.
\]

The KBGL gain over direct detection is then
\[
G_{\text{KBGL}} = \frac{P_{\text{success}}^{\text{KBGL}}}{P_{\text{success}}^{\text{direct}}} 
= \frac{\tau_{\text{meas}}}{\tau_{\text{meas}} + \tau_{\text{ent}}^0 e^{L/(2L_0)}} e^{L/(2L_0)}.
\]

This function grows exponentially until $L = 2L_0 \ln(\tau_{\text{meas}}/\tau_{\text{ent}}^0)$ (Fig.~\ref{SIfig_schemes_comparison_kbgl}c), after which it saturates to
\[
G_{\text{KBGL}}^{\text{max}} = \frac{\tau_{\text{meas}}}{\tau_{\text{ent}}^0}.
\]

The primary goal of a quantum memory-assisted telescope is then to maximize the ratio $\tau_{\text{meas}}/\tau_{\text{ent}}^0$. This can be achieved in several ways. One approach is to increase the entanglement generation rate by using short (i.e., broadband) photon pulses while detecting narrowband sources. For example, a 1~MHz zero-distance entanglement rate paired with a 1~kHz linewidth signal (e.g., from a laser) yields $G_{\text{KBGL}}^{\text{max}} = 10^3$, assuming negligible additional delays and average photon number below one per measurement. However, in practice, the entanglement rate is limited by factors such as the electron spin readout time (approximately $10~\mu$s), and generally more broadband astronomical sources are of significant interest.

\begin{figure*}[ht]
    \centering
    \includegraphics[width=1\textwidth]{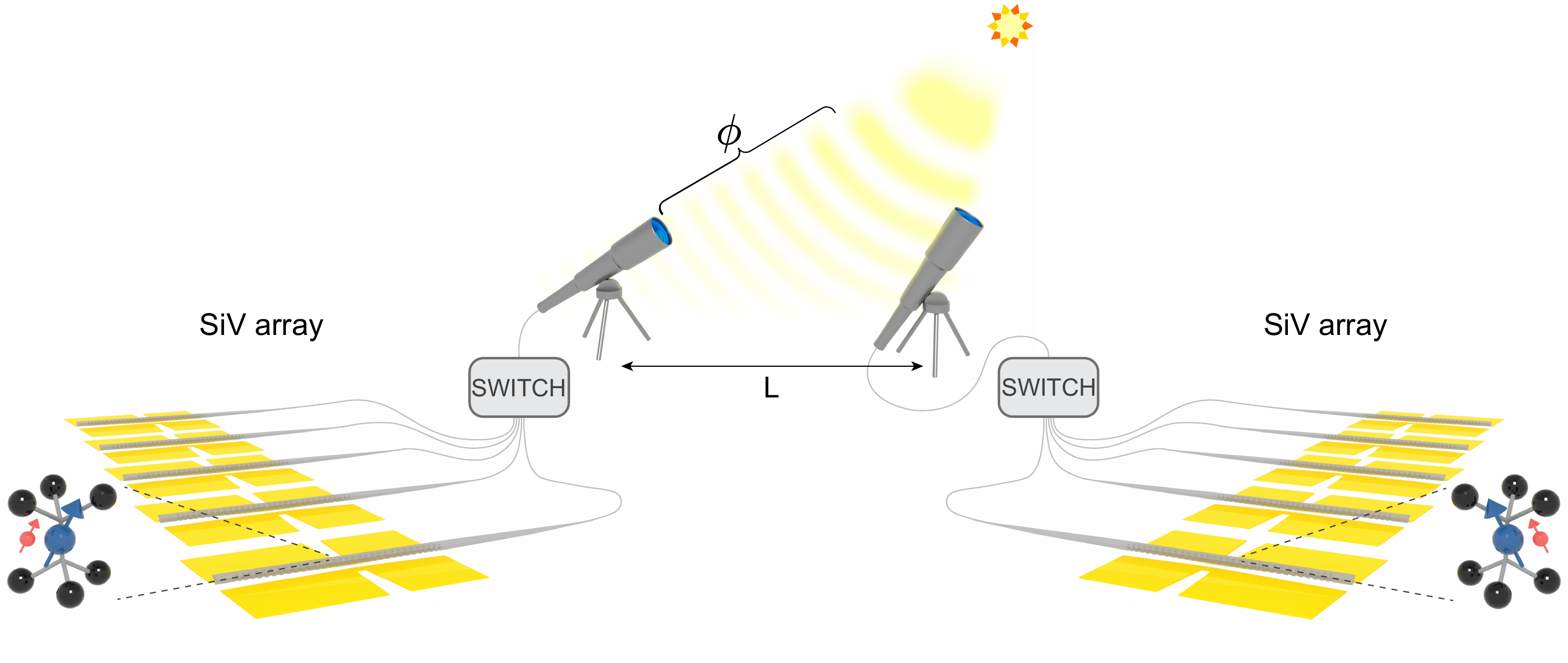}
    \caption{\textbf{Scalability of the KBGL scheme using on-chip SiV arrays and optical switches.}}
    \label{SIfig_multiplex}
\end{figure*}

Another strategy for enhancing gain is to use the multiplexing scheme proposed in~\cite{emil_telescopes}. The measurement window is divided into $N$ timebins corresponding to the coherence time of the signal photon (Fig.~\ref{SIfig_schemes_comparison_kbgl}a). Using binary encoding, only $\log_2(N)$ memory qubits per side are needed to store timing information. Then, $\log_2(N)$ Bell pairs enable a non-local parity check to identify the time bin with a photon without collapsing the memory qubits’ wavefunction. The memory qubits are subsequently measured in the $X$ basis to retrieve the signal light phase. This binary encoding significantly reduces overhead and extends the measurement window, provided the memory coherence time is sufficient.

When the measurement window exceeds the entanglement and processing time, exponential scaling is successfully avoided, and $P_{\text{success}}$ remains nearly constant with distance—unachievable in direct detection schemes. Taking the nuclear coherence time $T_2$ of \sivtn\ to be on the order of several seconds \cite{si29_node, two_fridge}, we use a 1-second measurement window in Fig.~\ref{SIfig_schemes_comparison_kbgl}b (dashed green). For $\Delta f = 1$~MHz signal light, this corresponds to approximately $10^6$~timebins, requiring $\log_2 (10^6)\approx20$~SiVs per interferometer node. We simulate this configuration to yield up to 3~orders of magnitude increase in success rate over direct detection (Fig.~\ref{SIfig_schemes_comparison_kbgl}c), without compromising Fisher information, unlike local measurement schemes such as homodyne and heterodyne detection. For the simulations, we assume telecom fiber transmission with a standard 0.3~dB/km attenuation and neglect frequency conversion losses, which would act as a constant multiplicative factor.

To reach this regime, two advances are necessary: a 1~kHz zero-distance entanglement rate (entanglement generation rate prior to including distance-dependent loss) and 20~SiVs per node. The former could be achieved with strain-induced SiV optical frequency tuning, eliminating losses incurred by frequency shifting. The latter can be realized using chips incorporating high-efficiency fiber packaging \cite{Zeng2023}. These findings highlight a clear pathway toward scalable, high-performance quantum-enhanced telescopes. This approach is particularly promising for detecting weak, narrowband signals. Moreover, periodic weak signals or signals with known arrival times could benefit from on-demand entanglement, where Bell pairs are distributed and stored in advance.

For relatively broadband signals ($\Delta f > \Delta f_\text{SiV} \approx 1$~GHz), one can employ the frequency-multiplexing scheme proposed in~\cite{emil_PRA}. In this approach, the total signal bandwidth $\Delta f$ is divided into smaller frequency windows, each of width $\Delta f_\text{SiV}$, corresponding to the bandwidth of a single SiV. The number of such windows is therefore $M = \Delta f / \Delta f_\text{SiV}$. Importantly, just as in the case of temporal multiplexing, the number of physical SiVs required to address these modes scales only logarithmically with $M$, namely $\log_2(M)$. This logarithmic scaling highlights a key advantage of the scheme: even for very broadband signals, the resource overhead in terms of SiV centers remains modest, enabling efficient and practical implementation of broadband quantum measurements.

\section{Photon erasure theory and simulations}

\begin{figure*}[ht]
    \centering
    \includegraphics[width=1\textwidth]{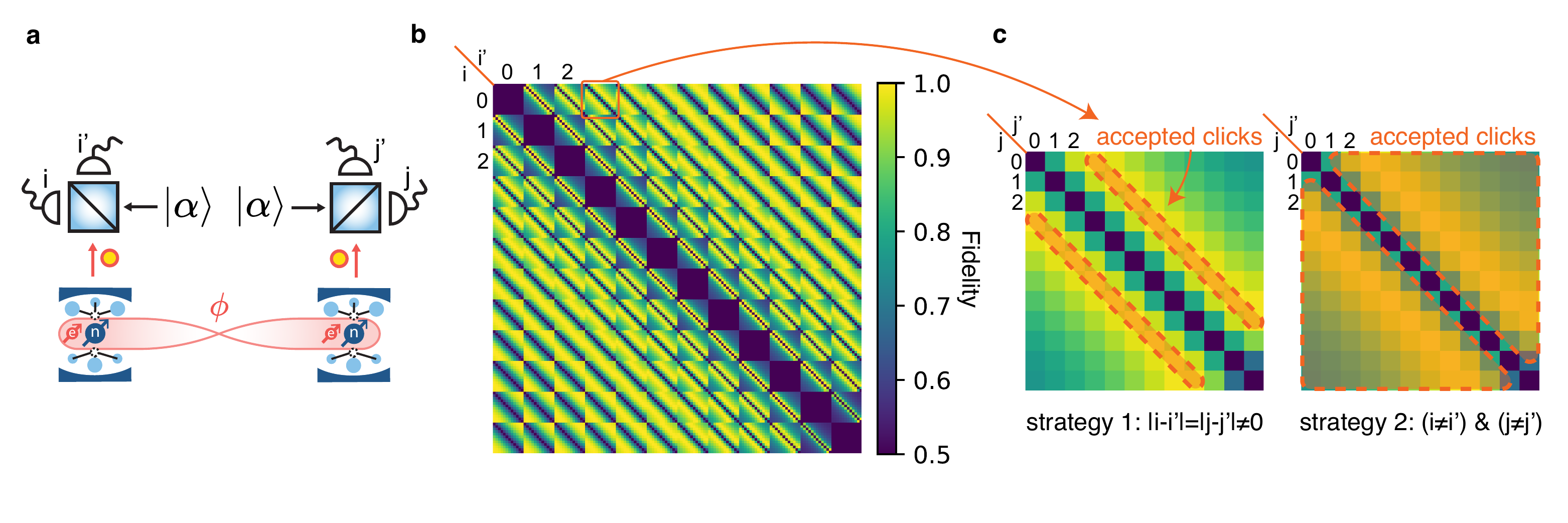}
    \caption{\textbf{Photon erasure strategies (no loss).} 
    a)~To preserve the Bell pair, which-path information must be erased. Signal light is combined with a weak coherent state on a beamsplitter. Detector click counts are $i$, $i'$ (left node) and $j$, $j'$ (right node). 
    b)~Simulated fidelity as a function of click configurations. Large squares represent \{$i$, $i'$\}, and small squares within show \{$j$, $j'$\}. 
    c)~Two erasure strategies: strategy 1 accepts outcomes with $|i-i'| = |j-j'| \neq 0$; strategy 2 accepts outcomes where $(i \neq i') \& (j \neq j')$. Orange outlines mark accepted events. Strategy 1 yields perfect fidelity in lossless conditions but with lower success probability.}
    \label{SIfig_erasure_perfect}
\end{figure*}

Photon erasure corresponds to an approximate $X$-basis measurement that removes the which-path information, i.e., whether a signal photon interacted with a particular node \cite{emil_PRA}. To achieve this, the reflected signal light is combined with a weak coherent state (local oscillator) on a beamsplitter, and specific detector click patterns are post-selected (Fig.~\ref{SIfig_erasure_perfect}a). We simulate the fidelity of the final nuclear spin state (expected to be $|\uparrow\downarrow \rangle + e^{i\phi} |\downarrow \uparrow\rangle$), given an incoming signal photon state $|10\rangle + e^{i\phi} |01\rangle$, a local oscillator state $|\alpha_{\text{LO}}\rangle$ at both nodes, and a specific detection outcome \{$i$, $i'$, $j$, $j'$\}. Here $\ket{01}$ ($\ket{\downarrow \uparrow}$) represents signal photon (spin states) at the right and left stations.

First, we consider the idealized situation without loss, MW errors, or dark counts. Before the beamsplitters and the interaction with SiVs, the state of the signal, LO, and spins can be written as $(\ket{10}+e^{i\phi}\ket{01})\ket{\alpha, \alpha, \downarrow, \downarrow}$ (up to normalization and assuming that the photon has already been non-locally heralded). After the interaction with the SiVs, the state becomes $(\ket{10}\ket{\uparrow\downarrow}+e^{i\phi}\ket{01}\ket{\downarrow\uparrow})\ket{\alpha, \alpha}$ or, equivalently, $(a^{\dagger}_{L}\sigma^X_L + e^{i\phi} a^{\dagger}_{R}\sigma^X_R)\ket{\alpha, \alpha,\downarrow,\downarrow}$, where $a^{\dagger}_{R}$ ($a^{\dagger}_{L}$) is the signal photon creation operator at the right (left) station, and $\sigma^X_R$ ($\sigma^X_L$) is the spin-flipping Pauli-$X$ operator at the right (left) station. After the beamsplitters, the state is 
\[
\ket{\Psi} = \{(a^{\dagger}_{L1}-a^{\dagger}_{L2})\sigma^X_L + e^{i\phi}(a^{\dagger}_{R1}-a^{\dagger}_{R2})\sigma^X_R\}\ket{\tfrac{\alpha}{\sqrt{2}}, \tfrac{\alpha}{\sqrt{2}}, \tfrac{\alpha}{\sqrt{2}}, \tfrac{\alpha}{\sqrt{2}},\downarrow,\downarrow},
\]
where indices $R_n$ ($L_n$), $n\in{\{1,2\}}$ denote detector paths at the right (left) station (detection outcome \{$i$, $i'$, $j$, $j'$\} on detectors \{$L1$, $L2$, $R1$, $R2$\}, Fig.~\ref{SIfig_erasure_perfect}a). Projecting this state onto a particular click event (omitting the common normalization factor $\tfrac{1}{2}e^{-2|\alpha|^2}$) yields:
\begin{align*}
\braket{i,i',j,j'|\Psi} &= \left[
\frac{\alpha^{i-1}\sqrt{i}}{\sqrt{(i-1)!}}
\frac{\alpha^{i'+j+j'}}{\sqrt{i'! j! j'!}}
-
\frac{\alpha^{i'-1}\sqrt{i'}}{\sqrt{(i'-1)!}}
\frac{\alpha^{i+j+j'}}{\sqrt{i! j! j'!}}
\right]\ket{\uparrow\downarrow} \\
&\quad + e^{i\phi}\left[
\frac{\alpha^{j-1}\sqrt{j}}{\sqrt{(j-1)!}}
\frac{\alpha^{i+i'+j'}}{\sqrt{i! i'! j'!}}
-
\frac{\alpha^{j'-1}\sqrt{j'}}{\sqrt{(j'-1)!}}
\frac{\alpha^{i+i'+j}}{\sqrt{i! i'! j!}}
\right]\ket{\downarrow\uparrow} \\[6pt]
&= \frac{\alpha^{i+i'+j+j'-1}}{\sqrt{i! i'! j! j'!}}
\left[(i-i')\ket{\uparrow\downarrow} + e^{i\phi}(j-j')\ket{\downarrow\uparrow}\right].
\end{align*}

From this expression, it is clear that the fidelity is perfect when $|i-i'| = |j-j'| \neq 0$ (a phase flip is required when $\text{sign}(i-i') \neq \text{sign}(j-j')$), and the fidelity is 50\% if $i=i'$ or $j=j'$ (Fig.~\ref{SIfig_erasure_perfect}b).

The number of detector clicks is denoted by $i$ and $i'$ for the left pair of detectors, and by $j$ and $j'$ for the right pair (Fig.~\ref{SIfig_erasure_perfect}a). 
Different click pattern combinations \{$i$, $i'$, $j$, $j'$\} yield different fidelities and success probabilities. 
This four-dimensional array of outcomes can be visualized as a two-dimensional matrix for clarity (Fig.~\ref{SIfig_erasure_perfect}b): 
each large square corresponds to a pair \{$i$,~$i'$\}, and the smaller squares within represent the corresponding values of \{$j$,~$j'$\}.

To structure these outcomes, we define two photon erasure strategies employed in this work:
\begin{itemize}
    \item \textbf{Strategy 1:} Accept only outcomes where $|i - i'| = |j - j'| \neq 0$.
    \item \textbf{Strategy 2 (used in Fig.~\ref{fig3}, \ref{fig4}, and \ref{fig5}):} Accept outcomes where $(i \neq i')$ and $(j \neq j')$.
\end{itemize}
Both strategies are illustrated in Fig.~\ref{SIfig_erasure_perfect}c. 
Although strategy~1 achieves perfect fidelity in the ideal lossless case, its success probability is lower than that of strategy~2 (Fig.~\ref{SIfig_erasure_loss}a).

The fidelity and efficiency of each strategy depend on the local oscillator amplitude $|\alpha_{\text{LO}}|$, but not on the intensity of the signal light being erased. While the fidelity map in Fig.~\ref{SIfig_erasure_perfect}b is independent of $|\alpha_{\text{LO}}|$, the probability of each click pattern is not, leading to $|\alpha_{\text{LO}}|$-dependent strategy performance.
\\

Incorporating experimental imperfections, three primary factors reduce the fidelity of photon erasure:

\begin{figure*}[ht]
    \centering
    \includegraphics[width=1\textwidth]{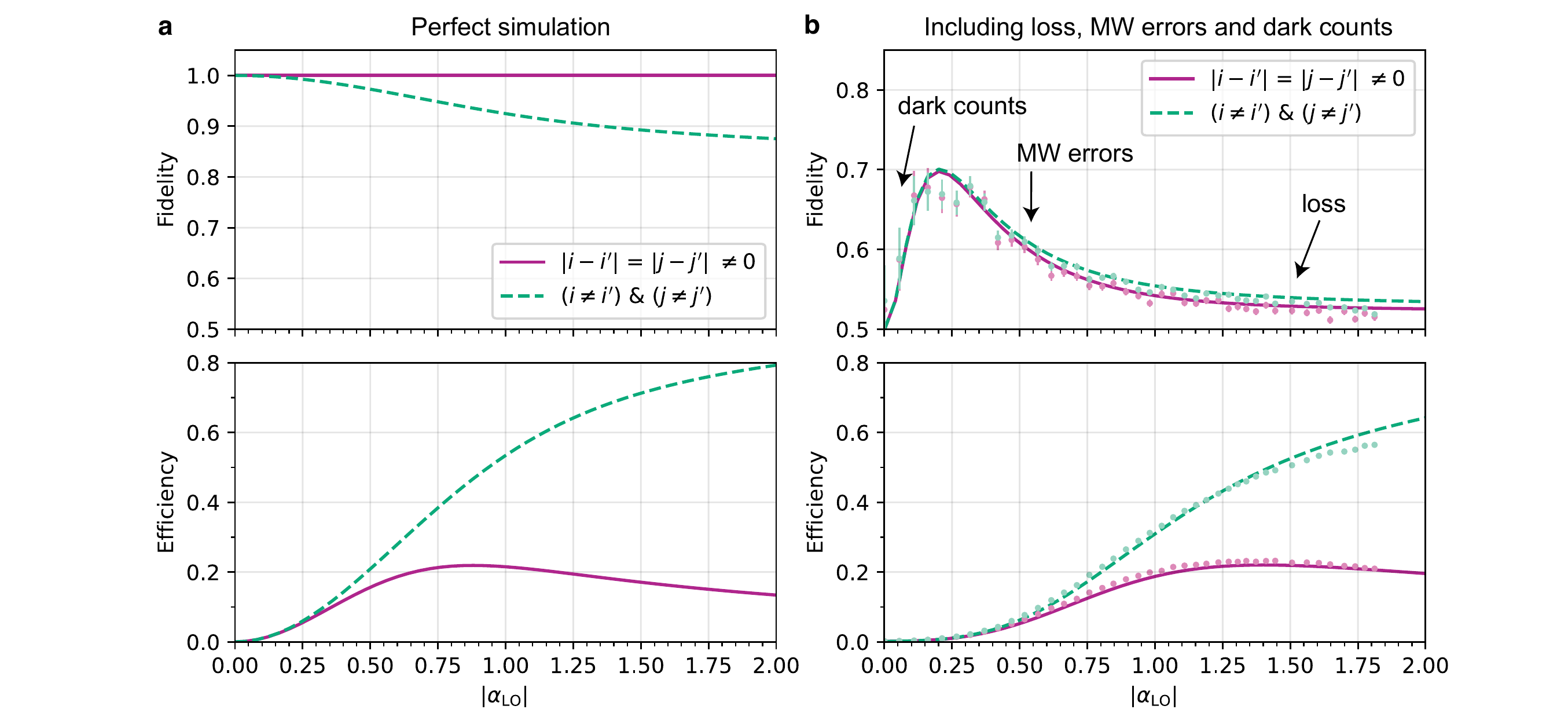}
    \caption{\textbf{Effect of imperfections on photon erasure strategies.} 
    a)~Ideal simulation: fidelity (top) and efficiency (bottom). 
    b)~Simulation with photon loss, MW errors, and dark counts. With pink and light green points corresponding to experimental data taken with strategy 1 and 2, respectively.
    Solid purple: strategy 1 ($|i-i'| = |j-j'| \neq 0$); dashed green: strategy 2 ($(i \neq i') \& (j \neq j')$).}
    \label{SIfig_erasure_loss}
\end{figure*}

\begin{enumerate}
    \item \textbf{Photon loss between the SiV and the detector.} 
    This is modeled as a fictitious beamsplitter that mixes the signal with a vacuum environment mode, which is later traced out of the final state.
    In the simulation shown in Fig.~\ref{SIfig_erasure_loss}b, we assume a total transmission of 15\% from SiV to detector. 
    Approximately 50\% of the loss is attributed to the amplitude-based reflection gate, while the remainder results from cavity-fiber coupling inefficiencies and detection setup inefficiencies. 
    Photon loss reduces fidelity, especially at high values of $|\alpha_{\text{LO}}|$, and shifts the efficiency peak toward higher amplitudes.

    \item \textbf{Microwave (MW) driving errors of the electron spin.} 
    Microwave and state preparation errors ($\varepsilon_{\text{MW}}$) lead to mis-heralding of the photon and lower state fidelity, and are modeled as a depolarizing channel. The probability of mis-heralding given a heralding event (correct or false) is
    \[
    p_{\text{mh}} = \frac{\varepsilon_{\text{MW}}}{\eta_{\mathrm{herald}}\mu_{\mathrm{sig}}+\varepsilon_{\text{MW}}},
    \]
    where $\eta_{\mathrm{herald}}$ is the heralding efficiency (50\% due to amplitude-based SMSPG) and $\mu_{\mathrm{sig}}$ is the signal mean photon number. State preparation errors $\varepsilon_{\text{MW}}$ also lower the fidelity of the initial state, resulting in constant factor, which lowers the fidelity of erasure.
    Then the resulting density matrix is: 
    \[
    \rho' = (1 - p_{\text{mh}}-\varepsilon_{\text{MW}})\rho + (p_{\text{mh}}+\varepsilon_{\text{MW}}) \frac{I}{2},
    \]
    where $I$ is the identity matrix. This process decreases the fidelity independently of $|\alpha_{\text{LO}}|$ and has no effect on the success probability. 
    We use $\varepsilon_{\text{MW}} = 6\%$, $\mu_{\mathrm{sig}}=0.15$, giving $p_{\text{mh}}=0.5$ in the simulation shown in Fig.~\ref{SIfig_erasure_loss}b.

    \item \textbf{Detector dark counts.} 
    These limit the maximum achievable fidelity in the low-$|\alpha_{\text{LO}}|$ regime. 
    We model this by assigning a small probability that the number of clicks on one detector increases by one, which is valid in the low-dark-count regime in which we operate.
    The total probability for any one of the four detectors to register a dark count is set to $p_{\text{DC}}/4 = 1 \times 10^{-2}$.
\end{enumerate}

Overall, while strategy~1 offers higher fidelity in the ideal loss-less and error-less case, strategy~2 proves more favorable under realistic conditions. By accounting for all these effects, we successfully reproduce the experimentally observed erasure fidelity and success probability (see Fig.~\ref{fig3}d,e).

\section{Efficiency and error breakdown}
\subsection{Entanglement rate and overall efficiency}
The efficiency of the overall phase probing protocol is mostly limited by the entanglement efficiency. The entanglement efficiency (detailed including repetition rates in Tab. \ref{SItable:rates}) is reduced by four major loss categories. The first one is component losses, including fiber coupling, circulator, and free-space AOM switch efficiency, as well as average cavity reflectivity. These losses are not fundamental and can be overcome with engineering improvements, such as for example fiber gluing \cite{Zeng2023} or efficient fast switches \cite{PsiQuantum2025}. The second major source of loss is frequency shifting through sideband generation using an EOM \cite{two_fridge, BQC_SiV}. This loss can be reduced by using a lower insertion-loss EOM as well as more efficient sideband generation through serrodyning. Alternatively, strain-induced SiV frequency tuning would bypass the need for frequency shifting. Third is the entangling protocol efficiency, due to the use of reflection amplitude-based spin-photon gates. Using reflection phase-based gates instead, the entangling protocol efficiency can in principle reach 1. The final limiting element is the average photon number $\mu$ of the WCS used for the entanglement. This could be improved by using a single photon source instead of a WCS.

Once entanglement has been heralded, the success probability of the phase probing measurement is the product of the heralding success probability, which depends on the average signal photon number as $\sim\mu_{\mathrm{sig}}$, and the erasure success probability depending on the LO strength and erasure strategy used (Fig.~\ref{fig3}d).
\begin{table}
	\begin{center}
		\begin{tabular}{c c c}
              & \multicolumn{2}{c}{Efficiency} \\
             \hline
              Fiber coupling & \multicolumn{2}{c}{70\%} \\
              Circulator & \multicolumn{2}{c}{70\%} \\
              Electron $\ket{\uparrow}$ cavity reflectivity & \multicolumn{2}{c}{80\%} \\
              Free-space AOM switch & \multicolumn{2}{c}{50\%} \\
              Frequency shifting \& filtering & \multicolumn{2}{c}{7.4\%} \\
              SNSPD & \multicolumn{2}{c}{95\%} \\
            
             \textbf{Total link efficiency} & \multicolumn{2}{c}{\textbf{1.5\%}} \\
             \hline
             WCS $\mu$ & \multicolumn{2}{c}{0.1-1} \\
             Entangling protocol efficiency  & \multicolumn{2}{c}{12.5\%} \\
             \hline
             \textbf{Total success probability} & \multicolumn{2}{c}{1.4 $\cdot 10^{-4}$-1.4 $\cdot 10^{-3}$} \\
             \hline
             \hline
             Repetition rate & 10 kHz (electron) & 2 kHz (nucleus) \\
             Duty cycle & \multicolumn{2}{c}{0.8} \\
             \hline 
             \hline
             \textbf{Entanglement rate} & 1.38 - 13.8 Hz (electron) & 0.28 - 2.8 Hz (nucleus)
        \end{tabular}
	\end{center}
	\caption{\textbf{Estimation of entanglement success probability and rate for the parallel entanglement scheme.} The total link efficiency describes the full insertion loss of our setup. The total success probability includes the total link efficiency as well as protocol-specific efficiencies.}
	\label{SItable:rates}
\end{table}

\subsection{Errors}

\begin{table}
	\begin{center}
		\begin{tabular}{c c c}
              & $e-e$ error & $n-n$ error \\
             \hline
              Initialization & $\sim$ 0 & 3\% \\
              MW gate error & 1\% & $\sim$ 0 (error detection) \\
              SiV optical contrast & \multicolumn{2}{c}{$\sim$ 0} \\
              Interferometer lock (visibility: 0.93) & \multicolumn{2}{c}{15\%} \\
              Readout & $\sim$ 0 & 5\% \\
              multi-photon error ($\mu \sim 0.1$) & \multicolumn{2}{c}{5\%} \\
            \hline
            \hline
             Estimated error & 19\% & 23\% \\
             \hline
             Measured error & 17.1\%(3.3\%) & 27\%(3.6\%)
        \end{tabular}
	\end{center}
	\caption{\textbf{Estimation of entanglement errors for the parallel entanglement scheme.} $\mu$: mean photon number per entanglement photonic qubit WCS. Uncertainties for measured errors are one standard deviation.}
	\label{SItable:errors}
\end{table}

The largest source of errors in the entanglement fidelity is the interferometer lock error, since even a small deviation $\delta$ in entanglement interferometer phase results in a relatively large reduction in entanglement fidelity $-5\delta^2$ (from expanding Eq. \ref{eq:final_ent_state} with $\delta\phi_e=\pi+\delta$ for small $\delta$). Further errors come from multi-photon events due to using a WCS as the entanglement photonic qubit, and MW gate errors. Nucleus-nucleus entanglement is not affected by MW gate errors during the entanglement operation itself because of error detection, but suffers from initialization and readout errors due to the additional steps required for these operations on the nuclear spin. \cite{si29_node, two_fridge, BQC_SiV} Similar to serial entanglement generation \cite{two_fridge}, heralding on $\ket{-}$ (as opposed to $\ket{+}$) photonic measurements for $\ket{\Psi^-}$ preparation cancels optical contrast to first order. Error estimations for entanglement are summarized in Tab. \ref{SItable:errors}.

Once nucleus-nucleus entanglement has been generated, the overall operation fidelity is further significantly reduced by the photon erasure step, detailed in the section: photon erasure theory and simulations.

\section{Interferometer phase locking }
\subsection{Phase locking protocol for entanglement generation}
\begin{figure*}
    \centering
    \includegraphics[width=1\textwidth]{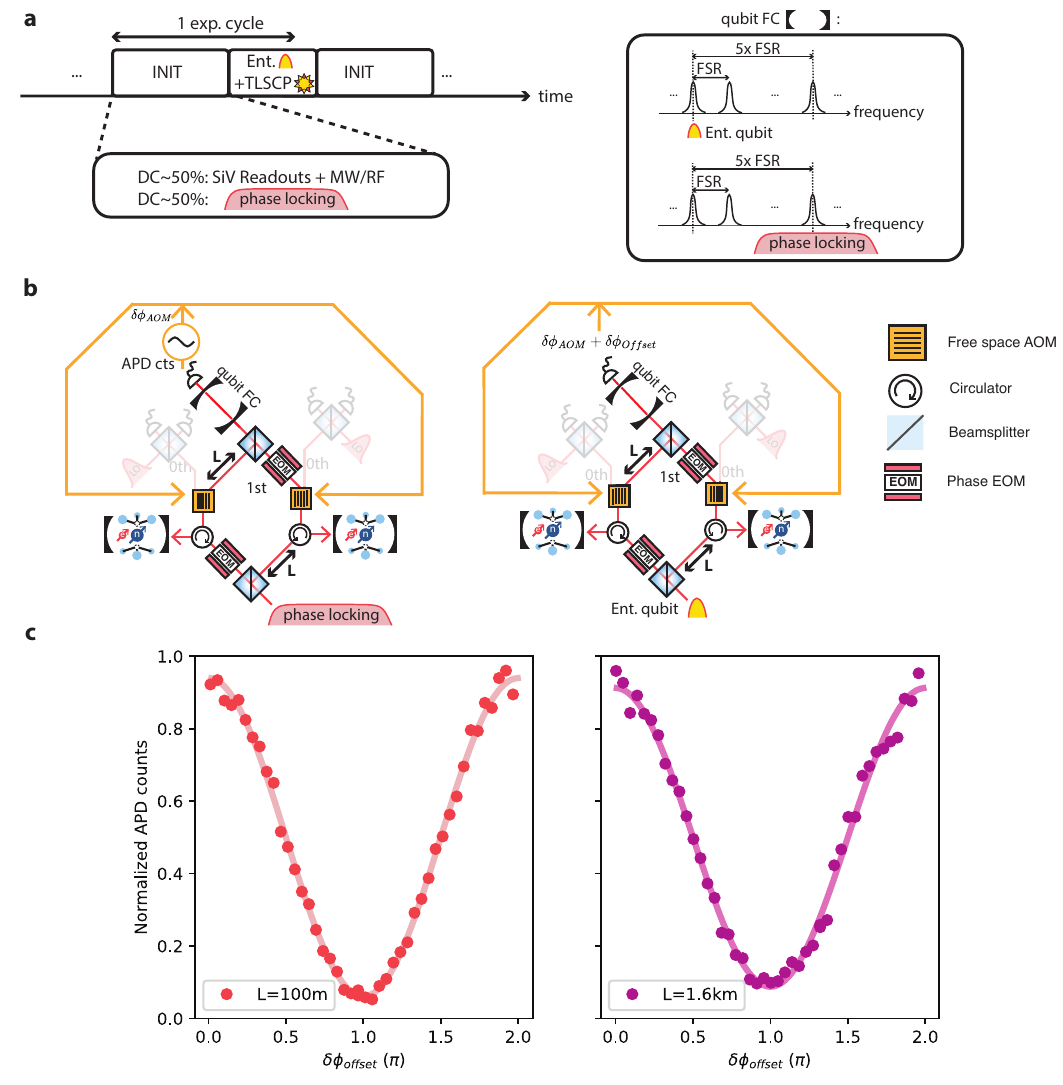}
    \caption{\textbf{Phase locking protocol.} a) Left, we interleave the SiV initialization step with phase locking. We adjust the duty cycle of the phase locking stage to be $\sim 50\%$ during the initialization procedure. Once the SiV is initialized, we attempt entanglement (Ent.) and a telescope measurement (TLSCP). Right, the frequency of the phase locking laser is put $5$ free spectral ranges (FSRs) of the qubit Fabry-Perot cavity away from the entanglement photonic qubit frequency. b) Left, during the phase locking stage, we use the avalanche photodiode (APD) photon counts as the input of the FPGA to update the RF phase ($\delta \phi_{\text{AOM}}$), which is split and fed back into two AOMs in the two stations. Right, during the entanglement and telescope stage, we drive  AOMs at both stations with the phase of RF signal  $\delta \phi_{\text{AOM}} + \delta\phi_{\text{Offset}}$. c) The phase locking performance when we sweep over $\delta\phi_{\text{Offset}}$ for $L=50$m (left) and $L=1.55$km (right). The contrast of the sweep are $15.5$ and $10.4$ for the $L=50$m and $L=1.55$km interferometer, respectively.}
    \label{SIfig_phaselocking}
\end{figure*}
To generate entanglement between two stations, it is necessary to lock the optical phase between two stations. We achieve this by interleaving our initialization step with phase locking feedback (Fig.~\ref{SIfig_phaselocking}a left). We adjust the duty cycle of the phase locking stage to be $\sim 50\%$ during the initialization procedure (Fig.~\ref{SIfig_phaselocking}a and  Fig.~\ref{SIfig_tlscp_logic}b), while the electro-optics modulators (EOM) always keep on. During the phase locking stage, we switch on the strong phase locking laser and read the photon counts from the end avalanche photodiode (APD). The frequency of the phase locking laser is put $5$ times the free spectral range (FSR) of 46.24 GHz of the entanglement qubit Fabry-Perot filtering cavity away from the entanglement photonic qubit frequency (Fig.~\ref{SIfig_phaselocking}a right). The 5-FSR difference is chosen such that (i) the phase locking light can transmit through the qubit Fabry-Perot cavity and therefore experiences the same frequency shifting as the entanglement photonic qubit and (ii) the laser frequency is far away from SiVs' resonant transitions so that the nuclear spins are not affected during the phase-locking stage. The 5-FSR difference $\Delta\lambda = 0.42$ nm is also much smaller than the qubit laser frequency $\sim 737$ nm so that the residual error is small $0.42/737 \approx 6\cdot 10^{-4}$.

We then use the APD photon counts as the input of the FPGA to update the RF phase ($\delta \phi_{\text{AOM}}$), which is split and fed back into two AOMs in the two stations (Fig.~\ref{SIfig_phaselocking}b left). We lock the APD photon counts to the reference port count, of which level roughly matches the steepest slope point of the interference pattern of the APD counts. During the entanglement and telescope stage, we use the updated $\delta \phi_{\text{AOM}}$, with a constant offset $\delta\phi_{\text{Offset}}$, to imprint the phase on the entanglement photonic qubit. Fig.~\ref{SIfig_phaselocking}c shows the normalized APD counts when sweeping $\delta\phi_{\text{Offset}}$. We pick $\delta\phi_{\text{Offset}}$ to minimize the APD counts for our entanglement protocol, which corresponds to locking at $\delta\phi_e = \pi$ (see main text). In practice, $\delta\phi_{\text{Offset}}$ drifts on the time scale of $1-2$ days.

\subsection{Interferometer and LO phase calibration}

\begin{figure*}
    \centering
    \includegraphics[width=1\textwidth]{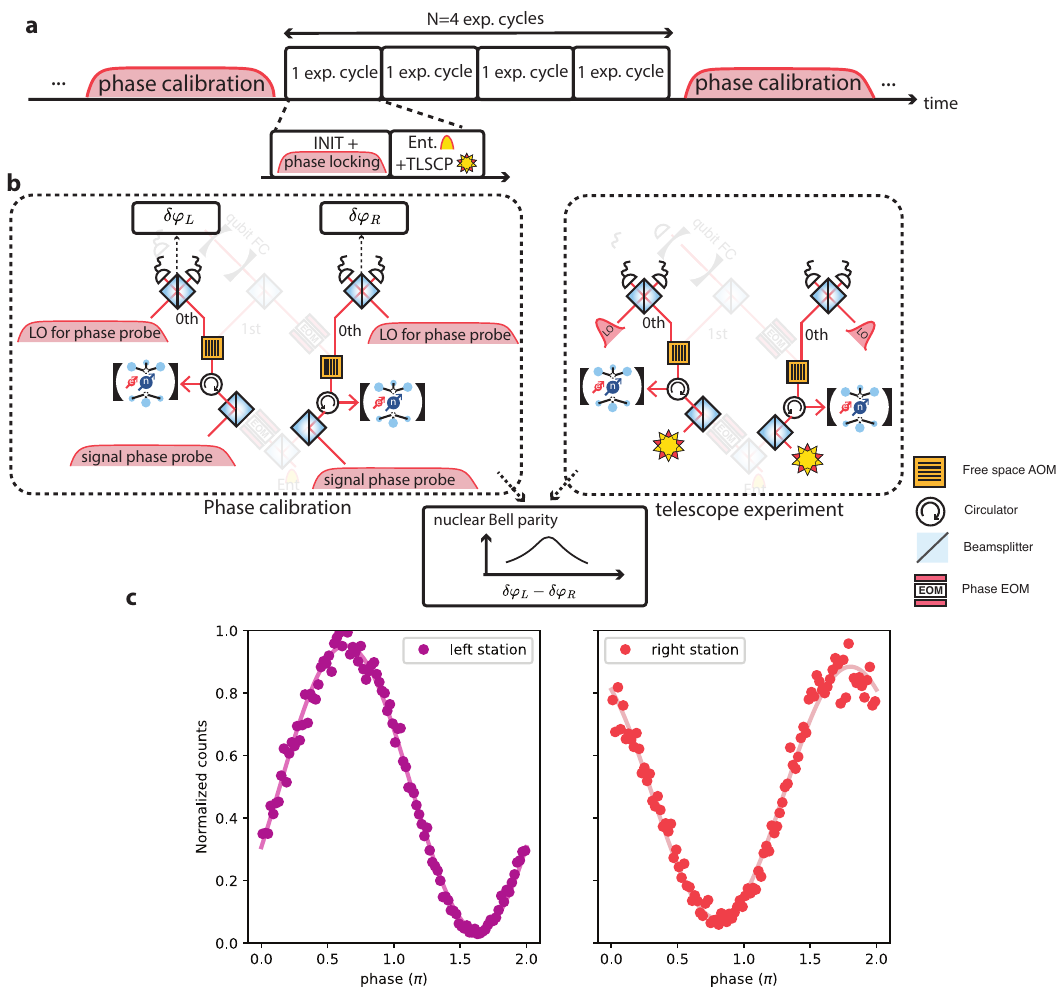}
    \caption{\textbf{Phase calibration for signal light} a) We probe the signal light phase between the two stations every $N = 4$ experiment cycles. b) In the phase probing stage, we send strong light and record the SNSPD photon counts in both nodes and determine the phase of two local signal-LO interferometers ($\delta\varphi_L$ and $\delta\varphi_R$). In the experiment stage, we switch off the phase probing light and run the sequence shown in the Fig.~\ref{fig3} in the main text, in which we extract the nuclear Bell-state parity conditioned on successful heralding events. We then correlate the nuclear Bell-state parity with the probed signal phase $\delta\varphi_L - \delta\varphi_R$. c) To check the phase probing performance, we independently send strong signal photon and correlate the detected photon counts with the probed phase. We measure the optical contrast to be $24.5$ ($11.5$) in the left (right) station.}
\label{SIfig_starphase}
\end{figure*}

In Fig.~\ref{fig4}, we do not actively lock and sweep the signal phase between the two stations but let the phase run freely and probe it periodically. Specifically, we probe the signal path phase between the two stations every $N=4$ runs of experiments (Fig.~\ref{SIfig_starphase}a). We do this by sending strong phase probing light for $82.5\mu$s and recording the SNSPD photon counts in both labs. We probe the phase twice, adding an extra $\pi/2$ phase on the local oscillator (LO) path using the fiber AOMs for the second probe. The two measurements together are then sufficient for us to determine the phase of two local signal-LO interferometer phases of each station ($\delta\varphi_L$ and $\delta\varphi_R$), as shown in Fig.~\ref{SIfig_starphase}b left (see Fig.~\ref{SIfig_setup} for details). For a nuclear $XX$ parity measurement the corresponding phase is then $\delta\varphi_L - \delta\varphi_R$ (Fig.~\ref{SIfig_starphase}b). 
To check the performance of the phase probing, we run a separate experiment in which we send strong signal light and record the SNSPD counts with the phase determined by phase probing, shown in Fig.~\ref{SIfig_starphase}c with an optical contrast of $24.5$ ($11.5$) at the left (right) station.

\section{Telescope measurement sequence logic}
\begin{figure*}
    \centering
    \includegraphics[width=1\textwidth]{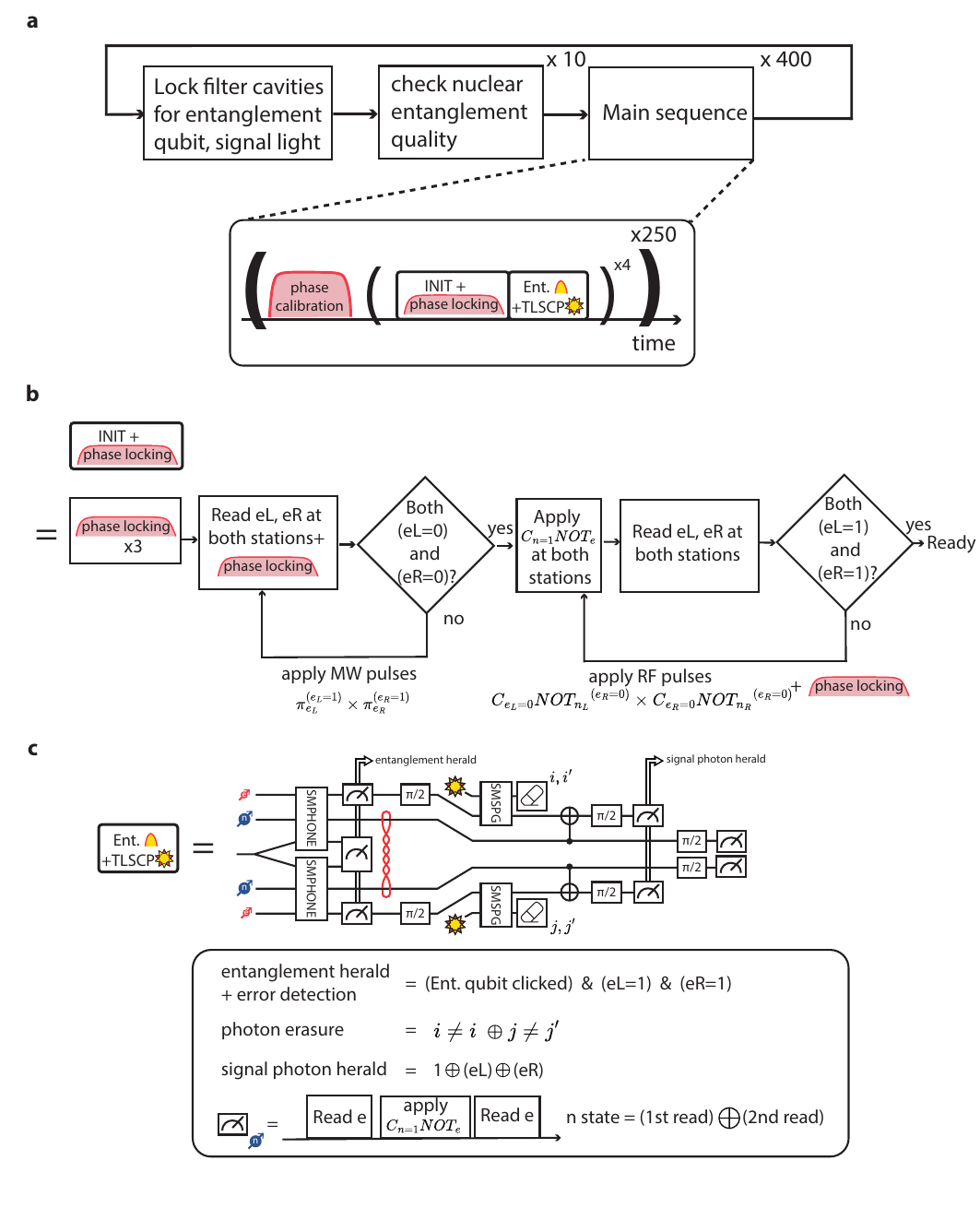}
    \caption{\textbf{a)} Flow chart of the experiments. Here, $e_{L(R)}$ denotes the electron spin state at the left (right) station, and $n_{L(R)}$ denotes the nuclear spin state at the left (right) station.\textbf{b)} Procedure for the initialization and phase locking. \textbf{c)} Procedure for the telescope sequence. }
    \label{SIfig_tlscp_logic}
\end{figure*}

Fig.~\ref{SIfig_tlscp_logic} outlines the experimental logic flowchart for the remote phase probing protocol. Fig.~\ref{SIfig_tlscp_logic}b details the SiV initialization protocol at the same time as the phase locking feedback, while Fig.~\ref{SIfig_tlscp_logic}c details the experimental sequence used in Fig.~\ref{fig4} and \ref{fig5} in the main text.

\section{Explicit state evolution with reflection amplitude-based gates}

The reflection based SMSPG relies on the electron spin state-dependent reflection amplitude: if the electron spin is the $\ket{\uparrow}(\ket{\downarrow})$ state, the incoming photon is reflected (transmitted or scattered). Because we do not have access to the transmission port of our devices, transmitted or scattered photons are lost. Therefore, describing the photonic state as $\ket{irl}$, where $i$ is the incoming photon mode, $r$ the reflected photon mode, and $l$ the lost photon mode, the SMSPG explicitly implements:

\begin{equation}
\ket{100}\ket{+_e}\longrightarrow (\ket{010}\ket{\uparrow_e} + \ket{001}\ket{\downarrow_e})/\sqrt{2}.
\end{equation}

During parallel entanglement generation, because entanglement is heralded by a click on the SNSPD after the final beamsplitter, this guarantees both that there was a single photon and that it was reflected (and thus not lost), so that the SMSPG effectively simplifies to $\ket{1_{\gamma}+_e}\longrightarrow\ket{1_{\gamma}\uparrow_e}$. This simplification breaks down for Fock states larger than 1, resulting in a reduced gate fidelity if the average photon number of the incoming WCS is too large.\cite{Nguyen2019PRL, Bhaskar2020, two_fridge}

For the remote phase sensing protocol, photon erasure implementation prevents signal photon heralding on photon detector clicks. We instead rely on the spin-photon interaction and subsequent spin measurement, which cannot distinguish between lost and reflected photons, effectively lowering the fidelity of the sequence. We note that additional photon losses between the spin-photon interaction and the photon detectors for erasure have the same fidelity-reducing effect. 

Starting with the two-station state right after telescope array arming:
$\ket{\psi} = \ket{+}_{e1}\ket{+}_{e2}\ket{\Psi^-}_{n1,n2}$, we denote the incoming photon state as $\ket{i_Li_Rr_Lr_Rl_Ll_R}$, where $i_{L(R)}$ is the incoming photon mode on the left (right) station, $r_{L(R)}$ the reflected photon mode from the left (right) station, and $l_{L(R)}$ is the lost photon mode of the left (right) station, so that:

\begin{multline}
(\ket{000000} + \sqrt{\mu}(\ket{100000}+e^{i\phi} \ket{010000})/\sqrt{2})\ket{+}_{e1}\ket{+}_{e2}\ket{\Psi^-}_{n1,n2} \\
\longrightarrow (\ket{000000}\ket{+}\ket{+} +\sqrt{\mu}/2(\ket{001000}\ket{\uparrow}\ket{+} +\ket{000010}\ket{\downarrow}\ket{+} + e^{i\phi} (\ket{000100}\ket{+}\ket{\uparrow} +\ket{000001}\ket{+}\ket{\downarrow})))\ket{\Psi^-}
\end{multline}

Applying the photon erasure step on the reflected photon modes (but not the lost photon modes) and taking the partial trace:

\begin{multline}
(\ket{+}\ket{+} + \sqrt{\mu}/2(\ket{\uparrow}\ket{+} + e^{i\phi}\ket{+}\ket{\uparrow}))(\bra{+}\bra{+} + \sqrt{\mu}/2(\bra{\uparrow}\bra{+} + e^{i\phi}\bra{+}\bra{\uparrow}))\ket{\Psi^-}\bra{\Psi^-} \\
+ \mu/4\ket{\downarrow}\bra{\downarrow}\ket{+}\bra{+}\ket{\Psi^-}\bra{\Psi^-} + \mu/4\ket{+}\bra{+}\ket{\downarrow}\bra{\downarrow}\ket{\Psi^-}\bra{\Psi^-}
\end{multline}

Applying C-$\pi$ gates and $\pi/2$ at each station:

\begin{multline}
(\ket{\uparrow}\ket{\downarrow}\ket{\uparrow\downarrow}-\ket{\downarrow}\ket{\uparrow}\ket{\downarrow\uparrow})/\sqrt{2} + \sqrt{\mu}/(2\sqrt{2})[(\ket{-}\ket{\downarrow}\ket{\uparrow\downarrow}-\ket{+}\ket{\uparrow}\ket{\downarrow\uparrow}) + e^{i\phi}(\ket{\uparrow}\ket{+}\ket{\uparrow\downarrow}-\ket{\downarrow}\ket{-}\ket{\downarrow\uparrow})])\bra{...} \\
+\mu/8(\ket{+}\ket{\downarrow}\ket{\uparrow\downarrow}-\ket{-}\ket{\uparrow}\ket{\downarrow\uparrow})\bra{...} +\mu/8(\ket{\uparrow}\ket{-}\ket{\uparrow\downarrow}-\ket{+}\ket{\downarrow}\ket{\downarrow\uparrow})\bra{...}.
\end{multline}

We then measure the electron qubits state and keep only the even parity results:

\begin{equation}
\ket{\uparrow}\ket{\uparrow}: \, (\ket{\uparrow\downarrow}-e^{-i\phi}\ket{\downarrow\uparrow})(\bra{\uparrow\downarrow}-e^{i\phi}\bra{\downarrow\uparrow})/4 + (\ket{\uparrow\downarrow}\bra{\uparrow\downarrow}+\ket{\downarrow\uparrow}\bra{\downarrow\uparrow})/4
\end{equation}
\begin{equation}
\ket{\downarrow}\ket{\downarrow}: \, (\ket{\uparrow\downarrow}-e^{i\phi}\ket{\downarrow\uparrow})(\bra{\uparrow\downarrow}-e^{-i\phi}\bra{\downarrow\uparrow})/4 + (\ket{\uparrow\downarrow}\bra{\uparrow\downarrow}+\ket{\downarrow\uparrow}\bra{\downarrow\uparrow})/4
\end{equation}

For both outcomes, the second term reduces the visibility of the phase $\phi$ measurement by half due to lost photons that are not erased.

\section{Improvement with phase-based spin-photon gates}
Higher cooperativity devices would enable the use of reflection phase-based gates instead of reflection amplitude-based gates for SMSPG. Here the photon is always reflected, but the photon incurs a phase shift depending on the electron spin state: $0$ if the electron is in the $\ket{\downarrow}$ state, $\pi$ if the electron is in the $\ket{\uparrow}$ state. This gate can then be described as $\ket{1+_e}\rightarrow\ket{1-_e}$. (i.e., this is simply a CZ gate.) With this gate, after the telescope array has been armed, the light signal is collected as follows:

\begin{multline}
(\ket{00} + \sqrt{\mu}(\ket{10}+e^{i\phi} \ket{01})/\sqrt{2})\ket{+}_{e1}\ket{+}_{e2}\ket{\Psi^-}_{n1,n2} \\
\longrightarrow (\ket{00}\ket{+}\ket{+} +\sqrt{\mu}/\sqrt{2}(\ket{10}\ket{-}\ket{+} +e^{i\phi}\ket{01}\ket{+}\ket{-}))\ket{\Psi^-}
\end{multline}

Erasing the photonic state:

\begin{equation}
(\ket{+}\ket{+} +\sqrt{\mu}/\sqrt{2}(\ket{-}\ket{+} +e^{i\phi}\ket{+}\ket{-}))\ket{\Psi^-},
\end{equation}

and then applying C-$\pi$ gates and $\pi/2$ at each station:

\begin{equation}
(\ket{\uparrow}\ket{\downarrow}\ket{\uparrow\downarrow}-\ket{\downarrow}\ket{\uparrow}\ket{\downarrow\uparrow})/\sqrt{2} + \sqrt{\mu}/2[(\ket{\downarrow}\ket{\downarrow}\ket{\uparrow\downarrow}-\ket{\uparrow}\ket{\uparrow}\ket{\downarrow\uparrow}) + e^{i\phi}(\ket{\uparrow}\ket{\uparrow}\ket{\uparrow\downarrow}-\ket{\downarrow}\ket{\downarrow}\ket{\downarrow\uparrow})]).
\end{equation}

Finally, we similarly measure the electron qubits state and keep only the even parity results:

\begin{equation}
\ket{\uparrow}\ket{\uparrow}: \, (\ket{\uparrow\downarrow}-e^{-i\phi}\ket{\downarrow\uparrow})/\sqrt{2}
\end{equation}
\begin{equation}
\ket{\downarrow}\ket{\downarrow}: \, (\ket{\uparrow\downarrow}-e^{i\phi}\ket{\downarrow\uparrow})/\sqrt{2}.
\end{equation}

Compared to the reflection amplitude-based gates, here there is no 50\% reduction in signal visibility due to lost photons. Furthermore, since the phase-based gate flips the electron to an orthogonal state - i.e., $\ket{+}\rightarrow\ket{-}$, $\langle+|-\rangle=0$ - (as opposed to the reflection amplitude-based gate: $\ket{+}\rightarrow\ket{\uparrow}$, $\langle+|\uparrow\rangle\neq0$), the non-local non-destructive photon heralding efficiency also improves by a factor of 2.

\section{Mis-heralding and overhead impact on signal-to-noise ratio scaling}

To improve the SNR of our non-local phase sensing protocol, we must address two sources of experimental imperfections. The first is experimental photonic loss overhead due to fiber coupling, optical circulator, and SMSPG efficiency. Photonic loss before the SiV reduces the efficiency ($p_{\mathrm{succ}}$), while photonic loss between the SiV and the erasure beamsplitter reduces the visibility ($V$). This results in a constant reduction of the SNR but does not affect its scaling with $\mu$.

The second experimental imperfection is the fidelity overhead associated with the pre-generated Bell state fidelity. This will reduce the visibility (once again causing a constant SNR reduction), but also introduces mis-heralding events. Mis-heralding causes the SNR scaling to ``curve down" from $\sqrt{\mu}$ (optimal scaling) to $\mu$ (local measurement scaling) for $\mu\lesssim \varepsilon_{\mathrm{mh}}/\eta_{\mathrm{herald}}$. 

\begin{figure*}[h]
    \centering
    \includegraphics[]{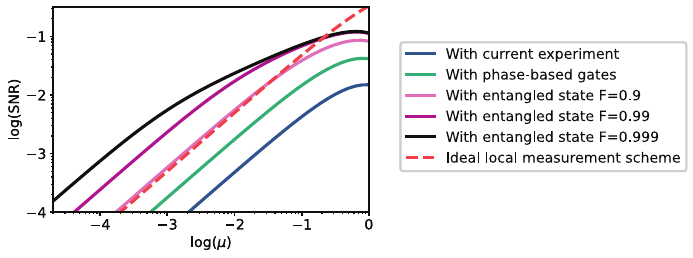}
    \caption{\textbf{Calculated SNR curves as a function of signal average photon number $\mu$ for different experimental parameters.} The blue curve uses current experimental parameters, the green curve is for replacing the amplitude-based SMSPG with phase-based gates, and the pink, purple, and black curves are with improved Bell state fidelities (still with phase-based gates). The dashed red curve is the SNR for an ideal local measurement scheme on a thermal photonic signal.}
    \label{SIfig_SNR_w_impr}
\end{figure*}

Replacing the amplitude-based SMSPG with phase-based gates can  increase the SNR by an approximately constant factor of 4 (Fig.~\ref{SIfig_SNR_w_impr}). Improving the Bell state fidelity both increases the SNR by a constant factor and increases the signal range for which the SNR scaling is optimal, resulting in a larger improvement over local measurement methods for smaller $\mu$. For Bell state fidelities above $\sim$90\%, the non-local phase sensing protocol can achieve a higher SNR than the ideal SNR of a local measurement method without any loss overhead of $\mathrm{SNR}=\mu/(2+\mu)$ \cite{Tsang2011}.

\end{document}